\documentclass[Journal]{IEEEtran}

\usepackage{./share/package_macros}
\newacronym{MP}{MP}{message passing}
\newacronym{MS}{MS}{\textit{min-sum}}
\newacronym{SP}{SP}{sum-product}
\newacronym{QoS}{QoS}{quality-of-service}
\newacronym{LUT}{LUT}{look-up table}
\newacronym{VN}{VN}{variable node}
\newacronym{CN}{CN}{check node}
\newacronym{DN}{DN}{decision node}

\newacronym{MAC}{MAC}{multiply-accumulate}
\newacronym{fsm}{FSM}{finite state machine}
\newacronym{snr}{SNR}{signal-to-noise ratio}
\newacronym{LLR}{LLR}{log likelihood ratio}
\newacronym{S/P}{S/P}{serial-to-parallel}
\newacronym{P/S}{P/S}{parallel-to-serial}

\newacronym{CAS}{CAS}{compare-and-select}

\newacronym{OMS}{OMS}{offset min-sum}

\begin{document} 
\title{A 588-Gbps LDPC Decoder Based on Finite-Alphabet Message Passing}


 \author{\IEEEauthorblockN{
 		Reza Ghanaatian,
 		Alexios Balatsoukas-Stimming,
 		Christoph M\"{u}ller,
 		Michael Meidlinger,
 		Gerald Matz,
 		Adam~Teman,
 		and Andreas Burg}
 		\thanks{R. Ghanaatian, A. Balatsoukas-Stimming, C. M\"{u}ller, A. Teman, and A. Burg are with the Telecommunications Circuits Laboratory, EPFL, Lausanne, Switzerland (email: \{reza.ghanaatian, alexios.balatsoukas, christoph.mueller, adam.teman, andreas.burg\}@epfl.ch).}
 		\thanks{M. Meidlinger and G. Matz are with the Vienna University of Technology, Vienna, Austria (email: \{mmeidlin, gmatz\}@nt.tuwien.ac.at).}
 } 

\maketitle

\begin{abstract}
An ultra-high throughput low-density parity check (LDPC) decoder with an unrolled full-parallel architecture is proposed,
which achieves the highest decoding throughput compared to previously reported LDPC decoders in the literature.
The decoder benefits from a serial message-transfer approach between the decoding stages 
to alleviate the well-known routing congestion problem in parallel LDPC decoders.
Furthermore, a finite-alphabet message passing algorithm is employed to replace the variable node update rule of the standard min-sum decoder with look-up tables, which are designed in a way that maximizes the mutual information between decoding messages.
The proposed algorithm results in an architecture with reduced bit-width messages, leading to a significantly higher decoding throughput and to a lower area as compared to a min-sum decoder when serial message-transfer is used.
The architecture is placed and routed for the standard min-sum reference decoder and for the proposed finite-alphabet decoder using a custom pseudo-hierarchical backend design strategy to further alleviate routing congestions and to handle the large design. 
Post-layout results show that the finite-alphabet decoder with the serial message-transfer architecture achieves a throughput as large as 588\,Gbps with an area of 16.2\,mm$^2$ and dissipates an average power of 22.7\,pJ per decoded bit in a 28\,nm~FD-SOI library.
Compared to the reference min-sum decoder, this corresponds to 3.1 times smaller area and 2 times better energy efficiency.

\end{abstract}


\IEEEpeerreviewmaketitle

\section{Introduction}\label{Intro}
\IEEEPARstart{L}{ow-density parity-check} (LDPC) codes have become the coding scheme of choice in high data-rate communication systems after their re-discovery in the 1990s~\cite{mackay1999good}, due to their excellent error correcting performance along with the availability of efficient high-throughput hardware implementations in modern CMOS technologies.
LDPC codes are commonly decoded using iterative \gls{MP} algorithms in which the initial estimations of the bits are improved by a continuous exchange of messages between decoder computation nodes.
Among the various \gls{MP} decoding algorithms, the \gls{MS} decoding algorithm~\cite{fossorier1999reduced} and its variants (e.g., offset \gls{MS}, scaled \gls{MS}) are the most common choices for hardware implementation. 
LDPC decoder hardware implementations traditionally start from one of these established algorithms (e.g., \gls{MS} decoding),
where the exchanged messages represent \glspl{LLR}.
These \glspl{LLR} are encoded as fixed point numbers in two's-complement or sign-magnitude representation, using a small number of uniform quantization levels, in order to realize the message update rules with low-complexity conventional arithmetic operations. 

Recently, there has been significant interest in the design of \emph{finite-alphabet} decoders for LDPC codes~\cite{planjery2013a, kurkoski2008a, balatsoukas2015sips, meidlinger2015asilomar, romero2015}. 
The main idea behind finite-alphabet LDPC decoders is to start 
from one or multiple arbitrary message alphabets, which can be encoded with a bit-width that is acceptable from an implementation complexity perspective.
The message update rules are then crafted as generic mapping functions to operate on this alphabet. 
The main advantage of such finite-alphabet decoders is that the message bit-width can be reduced significantly with respect to a conventional decoder, 
while maintaining the same error-correcting performance~\cite{balatsoukas2015sips,meidlinger2015asilomar}.
The downside of this approach is that the message update rules of finite-alphabet decoders usually cannot be described using fast and area-efficient standard arithmetic circuits.

Different hardware architectures for LDPC decoders have been proposed in the literature in order to fulfill the power and throughput requirements of various standards.
More specifically, various degrees of resource sharing result in flexible decoders with different area requirements.
On the one hand,~\emph{partial-parallel} LDPC decoders~\cite{zhang2010efficient, cevrero20105} and \textit{block-parallel} LDPC decoders~\cite{roth201015, kuo2008flexible} are designed for medium throughput, with modest silicon area.
\textit{Full-parallel}~\cite{blanksby2002690, onizawa2010design} and \emph{unrolled} LDPC decoders~\cite{schlafer2013new,balatsoukas2015sips}, on the other hand, achieve very high throughput (in the order of several tens or hundreds of Gbps) at the expense of large area requirements.
Even though, in principle, LDPC decoders are massively parallelizable, the implementation of ultra-high speed LDPC decoders still remains a challenge, especially for long LDPC codes with large node degrees~\cite{802.3an}.
While synthesis results for such \mbox{long-length} codes, as for example reported in~\cite{cushon34low}, show the potential for a very high throughput, the actual implementation requires several further considerations mainly due to severe routing problems and the impact of parasitic effects.

\subsubsection*{Contributions}
In this paper, we propose an \emph{unrolled full-parallel} architecture based on serial transfer of the decoding messages, which enables an ultra-high throughput implementation of LDPC decoders for codes with large node degrees by reducing the required interconnect wires for such decoders.
Moreover, we employ a finite-alphabet LDPC decoding algorithm in order to decrease the required quantization bit-width, and thus, to increase the throughput, which is limited by the serial message-transfer in the proposed architecture.
We also adopt a linear floorplan for the unrolled full-parallel architecture as well as an efficient pseudo-hierarchical flow that allow the high-speed physical implementation of the proposed decoder. 
To the best of our knowledge, by combining the aforementioned techniques, we present the fastest fully placed and routed LDPC decoder in the literature.

\subsubsection*{Outline}
The remainder of this paper is organized as follows:
Section~\ref{sec:Background} gives an introduction to decoding of LDPC codes, as well as more details on existing high-throughput implementations of LDPC decoders.
Section~\ref{sec:Decoder1} describes our proposed ultra-high throughout decoder architecture that employs a serial message-transfer technique.
In Section~\ref{sec:Decoder2}, our algorithm to design a finite-alphabet decoder with non-uniform quantization is explained and applied to the serial message-transfer decoder of Section~\ref{sec:Decoder1}.
Section~\ref{sec:Implementation} describes our proposed approach for the physical implementation and the timing and area optimization of our serial message-transfer decoders.
Finally, Section~\ref{sec:results} analyzes the implementation results, and Section~\ref{sec:conclusion} concludes the paper.

\section{Background}
\label{sec:Background}

In this section, we first briefly summarize the fundamentals of LDPC codes and the iterative MP algorithm for the decoding. We then review the state-of-the-art in high speed LDPC decoder architectures to set the stage for the description of our implementation.

\subsection{LDPC Codes and Decoding Algorithms}
A binary LDPC code is a set of codewords which are defined through an $M \times N$ binary-valued sparse parity check matrix as:
\begin{equation}
\big\lbrace \mathbf c \in \{0,1\}^N  \big | \mat H \mathbf c = \mathbf 0  \big\rbrace,
\end{equation}
where all operations are performed modulo~$2$. If the parity check matrix contains exactly $d_v$ ones per column and exactly $d_c$ ones per row, the code is called a $(d_v,d_c)$-regular LDPC code.
Such codes are usually represented with a \textit{Tanner graph}, which contains $N$ \glspl{VN} and $M$ \glspl{CN} and \gls{VN} $n$ is connected to \gls{CN} $m$ if and only if $\mat H _{mn}~=~1$.

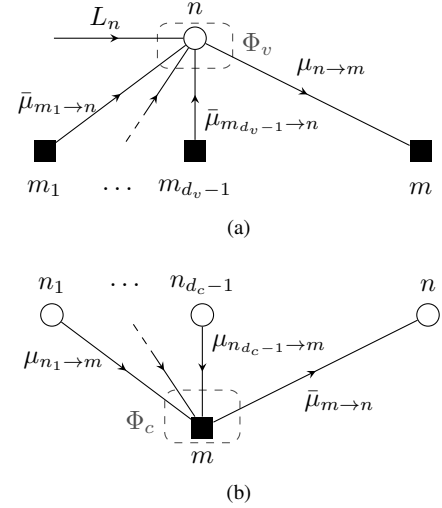
\begin{figure}%
	\centering
	\subfloat[][]{\begin{tikzpicture}[
   >=stealth,
   decoration={
    markings,
    mark=at position 0.5 with {\arrow{>}}}
    ] 

\tikzstyle{cnode}=[rectangle, inner sep = 4pt, fill=black]
\tikzstyle{vnode}=[draw=black, circle, inner sep =3pt]
   \draw (0   , 0) node[cnode] (m1) {};
   \draw (0   , -.5) node{$m_1$};

   \node (mphantom)  at (1   , 0) {};
   \draw (1   , -.5) node{$\dots$};

   \draw (2   , 0) node[cnode] (md) {};
   \draw (2   , -.5) node{$m_{d_v-1}$};

   \draw (5   , 0) node[cnode] (m) {};
   \draw (5   , -.5) node{$m$};

   \draw (2   , 1.5)   node[vnode] (n) {};
   \draw (2   , 1.9)  node{$n$};
   \draw[dashed, rounded corners=4pt, black!70]   (1.5   , 1.1) rectangle (2.5, 1.7) node[below right] {$\Phi_v$};

   \node (llr)  at (0   , 1.5) {};

   \draw[postaction={decorate}] (m1) -- (n) node[pos =.4, left]{$\bar\mu_{m_1\rightarrow n}$};
   \draw[postaction={decorate}, dash pattern=on 3pt off 3pt on 3pt off 3pt on 500pt] (mphantom) -- (n);
   \draw[postaction={decorate}] (md) -- (n) node[pos =.2, right]{$\bar\mu_{m_{d_v-1}\rightarrow n}$};
   \draw[postaction={decorate}] (n) -- (m) node[pos =.4, above right]{$\mu_{n\rightarrow m}$};

   \draw[postaction={decorate}] (llr) -- (n) node[pos =.4, above]{$L_n$};

\end{tikzpicture} \label{subfig:vnupdate} } \\
	\subfloat[][]{\begin{tikzpicture}[
   >=stealth,
decoration={
    markings,
    mark=at position 0.5 with {\arrow{>}}}
    ] 

\tikzstyle{cnode}=[rectangle, inner sep = 4pt, fill=black]
\tikzstyle{vnode}=[draw=black, circle, inner sep =3pt]
   \draw (0   , 1.5) node[vnode] (n1) {};
   \draw (0   , 1.9) node{$n_1$};

   \node (nphantom)  at (1   , 1.5) {};
   \draw (1   , 1.9) node{$\dots$};

   \draw (2   , 1.5) node[vnode] (nd) {};
   \draw (2   , 1.9) node{$n_{d_c-1}$};

   \draw (5   , 1.5) node[vnode] (n) {};
   \draw (5   , 1.9) node{$n$};

   \draw (2   , 0)   node[cnode] (m) {};
   \draw (2   , -.4)  node{$m$};
   \draw[dashed, rounded corners=4pt, black!70]   (2.5   , .5 ) rectangle (1.5, -.2) node[above left] {$\Phi_c$};

   \draw[postaction={decorate}] (n1) -- (m) node[pos =.4, left]{$\mu_{n_1\rightarrow m}$};
   \draw[postaction={decorate}, dash pattern=on 3pt off 3pt on 3pt off 3pt on 500pt] (nphantom) -- (m);
   \draw[postaction={decorate}] (nd) -- (m) node[pos =.2, right]{$\mu_{n_{d_c-1}\rightarrow m}$};
   \draw[postaction={decorate}] (m) -- (n) node[pos =.4, below right]{$\bar\mu_{m\rightarrow n}$};

\end{tikzpicture} \label{subfig:cnupdate} }
	\caption{
		\protect\subref{subfig:vnupdate} 
		\gls{VN} update 
		and \protect\subref{subfig:cnupdate} 
		\gls{CN} update 
		for $\set N (n)=\{m,m_1,\dots, m_{d_{v}-1}\}$ and $\set N (m)=\{n,n_1,\dots, n_{d_{c}-1}\}$.
	}%
	\label{fig:message_updates} 
\end{figure}

LDPC codes are traditionally decoded using \gls{MP} algorithms, where information is exchanged as messages between the \glspl{VN} and the \glspl{CN} over the course of several decoding iterations.
At each iteration the message from \gls{VN} $n$ to \gls{CN} $m$ is computed using a mapping 
$\Phi_v:  \mathbb{R}^{d_v}\rightarrow \mathbb{R}$, which is defined as:
\begin{align} \label{eqn:vnupdateGeneric}
\mu_{n\rightarrow m} = \Phi_v \big(L_n, \vec{\bar\mu}_{{\mathcal N}(n)\setminus m \rightarrow n} \big),
\end{align}
where $\set N(n)$ denotes the neighbors of node $n$ in the Tanner graph, $\vec{\bar\mu}_{{\mathcal N}(n)\setminus m \rightarrow n} \in \mathbb{R}^{d_v-1}$ 
is a vector that contains the incoming messages from all neighboring \glspl{CN} except $m$, and $L_n \in \mathbb{R}$ denotes the channel \gls{LLR} corresponding to \gls{VN}~$n$.
Similarly, the \gls{CN}-to-\gls{VN} messages are computed using a mapping $\Phi_c: \mathbb{R}^{d_c-1}\rightarrow \mathbb{R}$, which is defined as:
\begin{equation} \label{eqn:cnupdateGeneric}
\bar\mu_{m\rightarrow n} = \Phi_c \big(\vec{\mu}_{    {\mathcal N}(m)\setminus n \rightarrow m} \big).
\end{equation}
Fig.~\ref{fig:message_updates} illustrates the message updates in the Tanner graph. 
In addition to $\Phi_v$ and $\Phi_c$, a third mapping $\Phi_d: \mathbb{R}^{d_v+1}\rightarrow \{0,1\}$ 
is needed to provide an estimate of the transmitted codeword bits in the last \gls{VN} iteration based on the incoming 
\gls{CN} messages and the channel \gls{LLR} $L_n$ according to:
\begin{equation}\label{eqn:decupdate}
\hat c_n = \Phi_d (L_n, \vec{\bar \mu}_{\set N(n)\rightarrow n}).
\end{equation}
Messages are exchanged until a valid codeword has been decoded or until the maximum number of iterations $I$ has been reached. 

For the widely used \gls{MS} algorithm, the mappings \eqref{eqn:vnupdateGeneric} and \eqref{eqn:cnupdateGeneric} are:
	\begin{equation}\label{eqn:vnupdateMS}
	\Phi_v^{\mathrm{MS}}  (L, \vec{\bar\mu}  \big) = L + \sum_i \bar\mu_i,
	\end{equation}
and,
    \begin{equation}	\label{eqn:cnupdateMS}
	 \Phi_c^{\mathrm{MS}} (\vec \mu \big)
		= \sign \vec \mu \,\min |\vec \mu| ,
    \end{equation}
where $\min |\vec \mu|$ denotes the minimum of the absolute values of the vector components and $\sign \vec \mu = \prod_j \sign \mu_j$. 
The decision mapping $\Phi_d$ is defined as:
\begin{equation} \label{eqn:minsumDec}
\Phi_d^{\mathrm{MS}}(L,\vec{\bar\mu}) =  \frac{1}{2}\left(1-\sign \left (L + \sum_i \bar\mu_i \right)\right).
\end{equation}

\subsection{High Throughput LDPC Decoders}
Several high throughput LDPC decoders have been developed during the past decade in order to satisfy the high data-rate requirements of optical and high-speed Ethernet networks. These decoders usually rely on a full-parallel isomorphic~\cite{Hubert} architecture and a flooding schedule, which directly maps the algorithm for \emph{one} iteration to hardware. More specifically, the \gls{CN} and \gls{VN} update equations are directly mapped to $M$~\gls{CN} and $N$~\gls{VN} processing units and a hard-wired routing network is responsible for passing the messages between them.
From an implementation perspective, while such an architecture enables a very high throughput by fully exploiting the inherent parallelism of each iteration, the complexity of the highly unstructured routing network turns out to be a severe bottleneck.
In addition to this routing problem, such full-parallel decoders usually require one or two clock cycles for each iteration and in the worst case as many cycles as the maximum number of iterations for each codeword, which is another throughput limitation factor.

Several solutions have been proposed to alleviate the routing problem in full-parallel decoders, on both architectural and algorithmic levels. The authors of~\cite{darabiha2007interlaced,darabiha2008power} suggest using a bit-serial architecture, which only requires a single wire for each variable-to-check and check-to-variable node connection. While this approach can reduce the routing congestion, it also leads to a significant reduction in the decoding throughput. 
The decoder in~\cite{darabiha2008power}, for example, only achieves a throughput of $3.3$\,Gbps when implemented using a $130$\,nm CMOS technology.
Another architectural technique is reported in~\cite{onizawa2010design}, where the long wires of the decoder are partitioned into several short wires with pipeline registers. As a result, the critical path is broken down into shorter paths, but the decoder throughput is also affected since more cycles are required to accomplish each iteration. Nevertheless, the decoder of~\cite{onizawa2010design} is still able to achieve $13.2$\,Gbps in $90$\,nm CMOS with $16$ iterations.

On an algorithmic level, the authors of~\cite{mohsenin2010low} propose a \gls{MP} algorithm, called \textit{\gls{MS} split-row threshold}, which uses a column-wise division of the \mat{H} matrix into $S_{pn}$ partitions. Each partition contains $N/S_{pn}$ \glspl{VN} and $M$ \glspl{CN}, and global interconnects are minimized by only sharing the minimum signs between the \glspl{CN} of each partition. This algorithmic modification was used to implement a full-parallel decoder for the challenging $(2048, 1723)$ LDPC code in $65$\,nm CMOS, which achieves a throughput of $36.2$\,Gbps with $11$ decoding iterations.
Another decoder, reported in~\cite{mobini2009differential}, uses a hybrid hard/soft decoding algorithm, called \textit{differential binary \gls{MP}} algorithm, which reduces the interconnect complexity at the cost of some error-correcting performance degradation. A full-parallel $(2048, 1723)$ LDPC decoder using this algorithm was implemented in $65$\,nm CMOS technology,  achieving a throughput of up to $126$\,Gbps~\cite{cushon2014high}.
The work of \cite{cheng2014fully} also proposes another algorithmic level modification, called the probabilistic \gls{MS} algorithm, where a probabilistic second minimum value is used instead of the true second minimum value to simplify the \gls{CN} operation and to facilitate high-throughput implementation of full-parallel LDPC decoders. Further, a mix of tree and butterfly interconnect network is proposed in the CN unit to balance the interconnect complexity and the logic overhead and to reduce the routing complexity. The implementation of a decoder for the $(2048, 1723)$ LDPC code with the proposed techniques in $90$\,nm CMOS technology achieves a throughput of $45.42$\,Gbps.

Stochastic decoding of LDPC codes~\cite{tehrani2006stochastic} was another important improvement based on both algorithm and gate-level implementation considerations to solve the routing problem of LDPC decoders, where probabilities are interpreted as binary Bernoulli sequences. This approach, on the one hand, reduces the complexity of \glspl{CN} and the routing overhead, but, on the other hand, introduces difficulties in \gls{VN} update rules due to correlated stochastic streams, which may deteriorate the error-correcting performance especially in longer-length codes.
To solve this correlation problem by re-randomizing the \gls{VN} output streams, the work of~\cite{tehrani2010majority} proposes to use \textit{majority-based tracking forecast memories} in each \gls{VN}, which results in a decoder with full-parallel architecture for a $(2048, 1723)$ LDPC code that achieves a throughput of $61.3$\,Gbps in $90$\,nm CMOS.
An alternative method to track the probability values, called \textit{delayed stochastic decoding}, is reported in~\cite{naderi2011delayed} and the full-parallel decoder for the same code can deliver a throughput as large as $172.4$\,Gbps in $90$\,nm CMOS.

To solve the problem of throughput limitations in full-parallel decoders from potentially using multiple iterations for decoding, the work of~\cite{schlafer2013new} presents an unrolled full-parallel LDPC decoder.
In the proposed architecture, each decoding iteration is mapped to distinct hardware resources, leading to a decoder with $I$ iterations that can decode one codeword per clock cycle, at the cost of significantly increased area requirements with respect to non-unrolled full-parallel decoders.
This unrolled architecture achieves a throughput of $161$\,Gbps for a $(672, 546)$ LDPC code with $d_v=3$ and $d_c=6$, when implemented in a $65$\,nm CMOS technology.
It is noteworthy that an unrolled decoder has $50\%$ reduced wires between adjacent stages compared to a non-unrolled decoder since one stage of variable nodes is connected to one stage of check nodes with uni-directional data flow per decoding iteration.
Even though this measure leads to a lower routing congestion, it is still challenging to fully place and route such a decoder. 
This routing issue becomes more and more severe when considering longer LDPC codes and especially with increasing CN and VN degrees to achieve better error-correcting performance, as required in wireline applications such as for the $(2048, 1723)$ code with $d_v=6$ and $d_c=32$ used in the IEEE 802.3an standard~\cite{802.3an}.

\section{Serial Message-Transfer LDPC Decoder}
\label{sec:Decoder1}
\begin{figure*}
	\centering
	\includegraphics[width=7.0in]{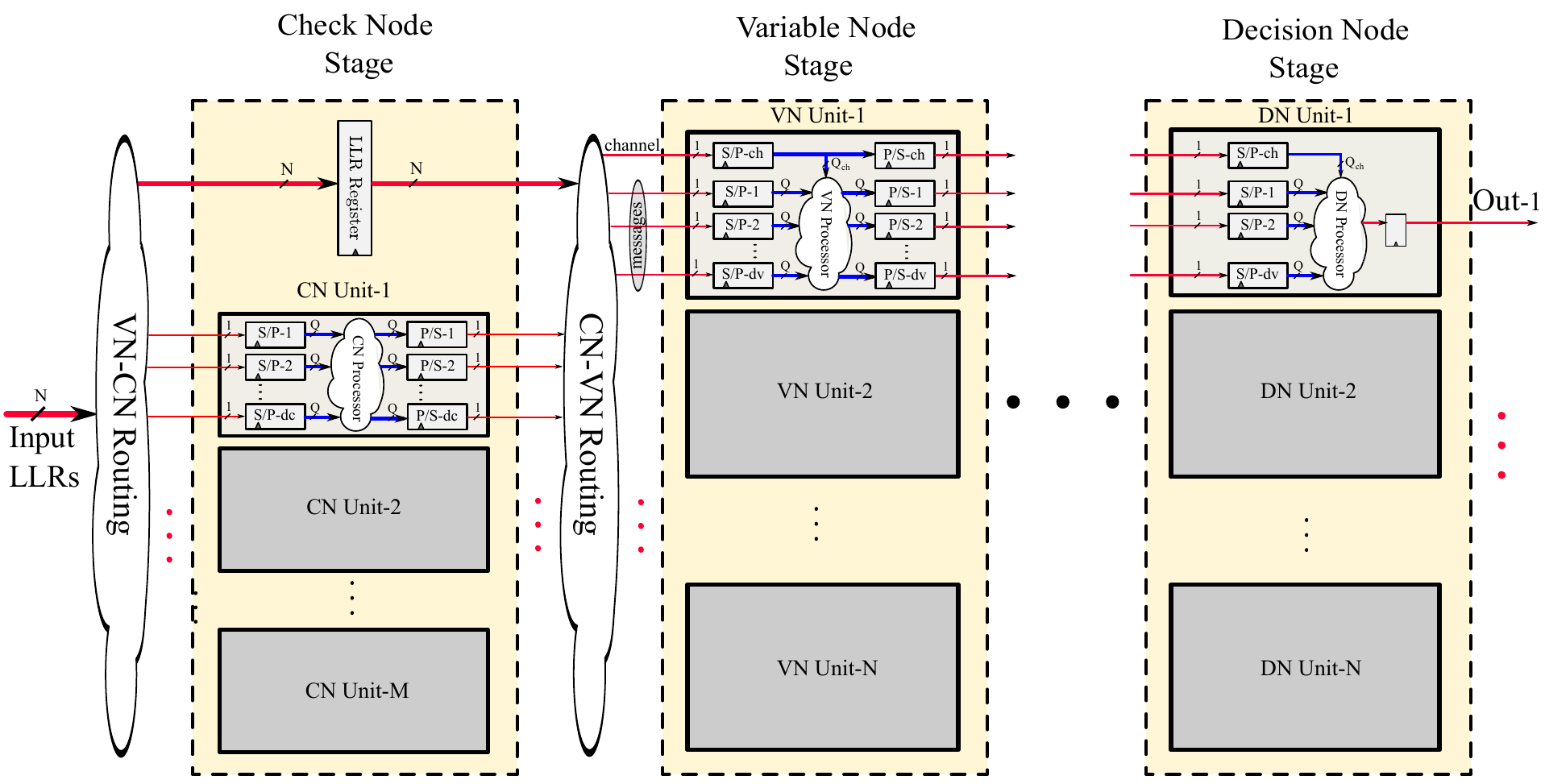}
	\caption{Serial message-transfer decoder architecture.}
	\vspace{-4mm}
	\label{fig:topleveldecoder}
\end{figure*}

Unrolled full-parallel LDPC decoders provide the ultimate throughput with smaller routing congestion than conventional full-parallel decoders.
However, they are still not trivial to implement for long LDPC codes with high \gls{CN} and \gls{VN} degrees, which suffer from severe routing congestion. 
Hence, in this section, we propose an unrolled full-parallel LDPC decoder architecture that employs a serial message-transfer technique between \glspl{CN} and \glspl{VN}.\footnote{We note that a first implementation of our decoder was based on parallel (word-level) message-transfer. The place and route tool for this implementation was hardly able to converge even when the area utilization was unacceptable. We, therefore, propose serial message-transfer architecture and we adopt special implementation methodologies for this architecture, which will be explained in Section~\ref{sec:Implementation}.} This architecture is similar to the bit-serial implementations of~\cite{darabiha2007interlaced,darabiha2008power} in the way the messages are transferred; however, as we shall see later, it differs in the fact that it is unrolled and in the way the messages are processed in the \glspl{CN} and \glspl{VN}.

\subsection{Decoder Architecture Overview}
An overview of the proposed unrolled serial message-transfer LDPC decoder architecture is shown in Fig.~\ref{fig:topleveldecoder}. As with all unrolled LDPC decoders, each decoding iteration is mapped to a distinct set of $N$ \gls{VN} and $M$ \gls{CN} units, which form a processing pipeline. In essence, the unrolled LDPC decoder is a systolic array, in which a new set of $N$ channel \glspl{LLR} is read in every clock cycle and a decoded codeword is put out in every clock cycle.

Even though both the \glspl{CN} and \glspl{VN} can compute their outgoing messages in a single clock cycle, similar to the architecture in~\cite{balatsoukas2015sips}, in the proposed serial message-transfer architecture each message is transfered one bit at a time between the consecutive variable node and check node stages over the course of $Q_{\text{msg}}$ clock cycles, where $Q_{\text{msg}}$ is the number of bits used for the messages.
More specifically, each \gls{CN} and \gls{VN} unit contains a \gls{S/P} and \gls{P/S} conversion unit at the input and output, respectively, which are clocked $Q_{\text{msg}}$ times faster than the processing clock to collect and transfer messages serially, while keeping the overall decoding throughput constant. More details on the architecture of the \gls{CN} and \gls{VN} units as well as the proposed serial message-transfer mechanism are provided in the sequel.

\subsection{Decoder Stages}
The unrolled LDPC decoder, illustrated in Fig.~\ref{fig:topleveldecoder}, consists of three types of processing stages, which are described in more detail below.
We note that the CN/VN processors of this reference decoder are similar to those of a standard \gls{MS} decoder, and our modifications for these parts (to realize a finite-alphabet decoder) are discussed in Section~\ref{sec:Decoder2}.

\subsubsection{Check Node Stage}
Each check node stage consists of $M$ \gls{CN} units, each of which contains three components: a \gls{CN} processor, which implements \eqref{eqn:cnupdateMS} similarly to \cite{balatsoukas2015sips, schlafer2013new}, $d_c$ \gls{S/P} units for the $d_c$ input messages, and $d_c$ \gls{P/S} units for the $d_c$ output messages. Moreover, the complete check node stage contains a register bank that is used to store the channel \glspl{LLR}, which are not directly needed by the check node stage, but nevertheless must be forwarded to the following variable node stage and thus need to be buffered. Hence, no \gls{S/P} and \gls{P/S} units are required for the channel \gls{LLR} buffers in the check node stages as they are simply forwarded serially to the following variable node stage.

\subsubsection{Variable Node Stage}
Each variable node stage consists of $N$ \gls{VN} units, each of which contains a \gls{VN} processor and \gls{S/P} and \gls{P/S} units at the inputs and outputs, respectively, similar to the \gls{CN} unit structure. Each \gls{VN} processor implements the update rule \eqref{eqn:vnupdateMS} similarly to \cite{balatsoukas2015sips}.

\subsubsection{Decision Node Stage}
The last variable node stage is called a decision node stage because it is responsible for taking the final hard decisions on the decoded codeword bits. The structure of this stage is similar to a variable node stage, but a \gls{DN} has a simpler version of the \gls{VN} processor that only computes sum of all inputs and put out its sign~\cite{balatsoukas2015sips}, and thus no \gls{P/S} unit is required at its output.

\subsection{Message Transfer Mechanism}
\begin{figure}
	\centering
	\includegraphics[width=3.5in]{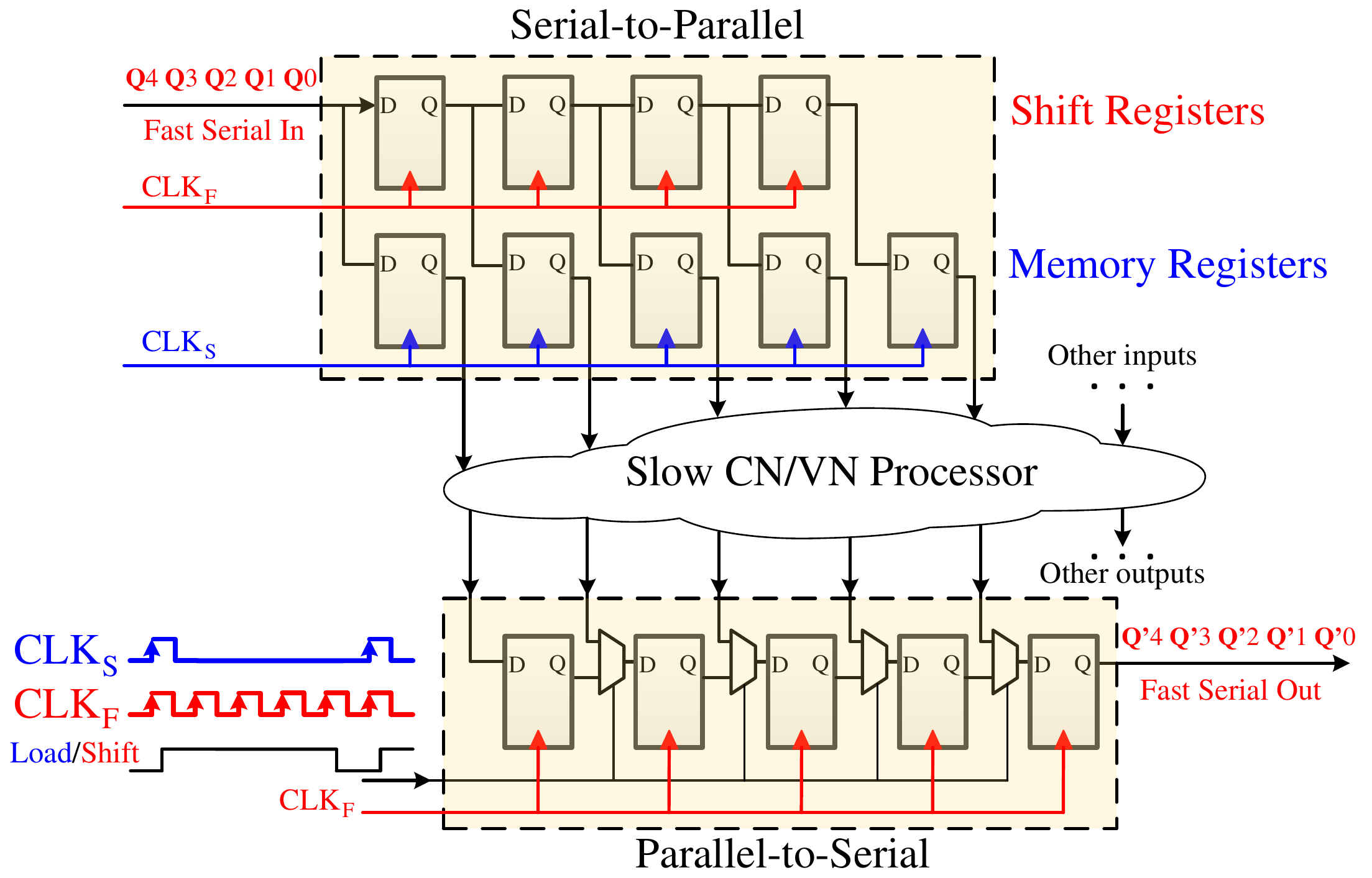}
	\caption{The message receive and transfer mechanism by \gls{S/P} and \gls{P/S} shift registers, enabled by the fast clock (\fastclk), and the message process enabled by the slow clock (\slowclk) for $Q_\text{msg}=5$.}
	\label{fig:shr}
\end{figure}

One of the modifications, compared to \cite{schlafer2013new} and \cite{balatsoukas2015sips}, 
is the serial transfer of the channel and message \glspl{LLR} between the stages of the decoder, which reduces the required routing resources by factor of $Q_{\text{msg}}$.
This modification is applied to make the placement and routing of the decoder feasible, especially for large values of $d_v$ and $d_c$. 
To this end, as explained in the previous section, a \gls{S/P} and a \gls{P/S} shift register are added to each input and each output of the \gls{CN} and \gls{VN} units, as illustrated in Fig.~\ref{fig:shr}.
We see that the \gls{S/P} unit consists of a ($Q_{\text{msg}}-$1)-bit shift register and $Q_{\text{msg}}$ memory registers, while the \gls{P/S} unit has $Q_{\text{msg}}$ registers with multiplexed inputs.
The serial messages are transfered with a fast clock, denoted by \fastclk, that is $Q_{\text{msg}}$ times faster than the slow processing clock, denoted by \slowclk. More specifically, at each \gls{CN} unit and \gls{VN} unit input, data is loaded serially into the \gls{S/P} shift register using the fast \fastclk, and after the $Q_{\text{msg}}$-th cycle all message bits are stored in memory registers, clocked by the slow \slowclk. The \gls{CN}/\gls{VN} processing can then be performed in one \slowclk cycle and the output messages are saved in the output \gls{P/S} shift register and transferred serially to the next stage using again \fastclk. At the same time, a new set of messages is serially loaded into the input shift register. We note that for simpler clock tree generation, all registers in Fig.~\ref{fig:shr} are clocked by \fastclk, while \slowclk is actually implemented as a pulsed clock, which is generated using a clock-gating cell controlled by a finite state machine.


\subsection{Decoder Hardware Complexity and Performance Analysis}
In this section, we describe the required memory complexity of the proposed decoder as well as the decoding latency and the throughput.

\subsubsection{Memory Requirement}
The memory complexity can easily be characterized by counting the number of required registers and can be approximated by:
\begin{equation} \label{eqn:finaltotalreg}
R_{\text{tot}}\approx N(d_v+1)Q(6I-1),
\end{equation}
where $I$ is the number of decoding iterations (which in unrolled decoders strongly affects the memory requirements) and \mbox{$Q=Q_{\text{msg}}=Q_{\text{ch}}$}, which is often the case for \gls{MS} LDPC decoders.
From \eqref{eqn:finaltotalreg}, one can easily see that the quantization bit-width linearly increases the memory requirement for the proposed architecture.

\subsubsection{Decoding Latency}
Since each stage has a delay of two \slowclk cycles and there are two stages for each decoding iteration, the decoder latency is $4I$ \slowclk cycles or, equivalently, $4IQ_\text{max}$ \fastclk cycles, where $Q_{\text{max}}=\max (Q_{\text{msg}},Q_{\text{ch}})$.

\subsubsection{Decoding Throughput}
In the proposed unrolled architecture, one decoder codeword is output in each \slowclk cycle. Therefore, the coded throughput of the decoder is: 
\begin{align}
	T=Nf_{\text{S}_\text{max}},
\end{align}
where $f_{\text{S}_\text{max}}$ is the maximum frequency of \slowclk while the maximum frequency of \fastclk or simply maximum frequency of the decoder is $f_{\text{max}}=f_{\text{F}_\text{max}}=Q_\text{max}f_{\text{S}_\text{max}}$.
For the proposed architecture, we have:
\begin{equation}\label{eq:fmax}
f_{\text{S}_\text{max}}=\big\{\max \big((Q_\text{max}T_{\text{CP},\text{route}}),\,(T_{\text{CP},\text{VN}}),\,(T_{\text{CP},\text{CN}})\big) \big\} ^{-1} , 
\end{equation}
where $T_{\text{CP},\text{VN}}$ and $T_{\text{CP},\text{CN}}$ are the delay of the critical paths of the \gls{CN} unit and the \gls{VN} unit, respectively, and $T_{\text{CP},\text{route}}$  is the critical path delay of the (serial) routing between the decoding stages.
Thus, the decoder throughput will be limited by the routing, if the VN/CN delay is smaller than $Q_\text{max}$ times the routing delay. Hence, on the one hand, the serial message-transfer decoder alleviates the routing problem by reducing the required number of wires, but on the other hand, the decoder throughput for large quantization bit-widths may be affected, as the serial message-transfer delay will become the limiting factor.

\section{Finite-Alphabet Serial Message-Transfer LDPC Decoder}\label{sec:Decoder2}
Even though the serial message-transfer architecture alleviates the routing congestion of an unrolled full-parallel LDPC decoder, it has a negative impact on both throughput and hardware complexity, as discussed in the previous section. In our previous work~\cite{meidlinger2015asilomar,balatsoukas2015sips}, we have shown that finite-alphabet decoders have the potential to reduce the required number of message bits while maintaining the same error rate performance. In this section, we will review the basic idea and our design method for this new type of decoders and then show how the bit-width reduction technique of~\cite{meidlinger2015asilomar,balatsoukas2015sips} can be applied verbatim in order to increase the throughput and reduce the area of the serial message-transfer architecture.

\subsection{Mutual Information Based Finite-Alphabet Decoder}
In our approach of~\cite{balatsoukas2015sips,meidlinger2015asilomar}, the standard message-passing decoding algorithm update rules are replaced by custom update rules that can be implemented as simple \glspl{LUT}. These \glspl{LUT} take integer-valued input messages and produce a corresponding output message. Moreover, the input-output mapping that is represented by the \glspl{LUT} is designed in a way that maximizes the mutual information between the \gls{LUT} output messages and the codeword bit that these messages correspond to. We note that a similar approach was also used in~\cite{romero2015}, but the corresponding hardware implementation would have a much higher hardware complexity than the method of~\cite{balatsoukas2015sips,meidlinger2015asilomar}. This happens because, contrary to~\cite{romero2015}, in~\cite{balatsoukas2015sips,meidlinger2015asilomar} we used \glspl{LUT} only for the \glspl{VN} while the \glspl{CN} use the standard min-sum update rule.

\subsection{Error-Correcting Performance and Bit-Width Reduction}
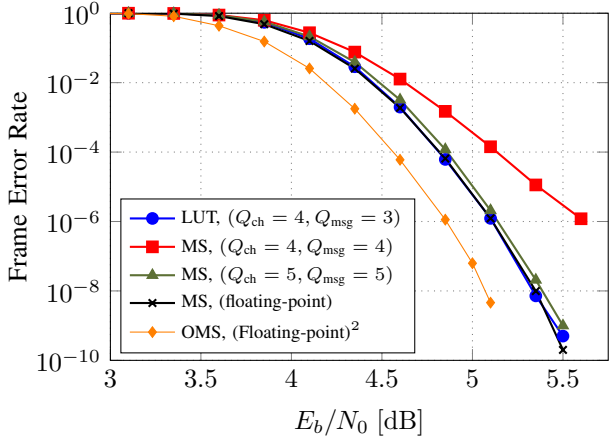
\begin{figure}[t]
	\centering
	\begin{tikzpicture}
\pgfplotsset{grid style={dotted,gray}}
\begin{semilogyaxis}[
    width=8.2cm,
    height=6.2cm,
    xlabel={$E_b/N_0\;[\mathrm{dB}]$},
    ylabel={Frame Error Rate},
    title={},
		ymin = 1e-10, ymax = 1e0,
		ytick = {1e0,1e-2,1e-4,1e-6,1e-8,1e-10},
		xmin = 3, xmax = 5.75,
		grid = both,
    legend style={
        cells={anchor=west}, 
        at={(.02,.02)},        
        anchor=south west    
    }
   ]

   \addplot+[
			thick,
      mark=*,
      mark options={fill=blue}, 
      color=blue,
      ] 
      table [x index             = 0,
             y  index            = 1, 
             col sep=comma] {figure/lut.csv};
   \addlegendentry{\scriptsize LUT, $(Q_{\text{ch}} = 4, Q_{\text{msg}} = 3)$}

   \addplot+[
			thick,
      mark=square*,
      mark options={fill=red},
      color = red,
      ] 
      table [x index             = 0,
             y  index            = 1, 
             col sep=comma] {figure/fixed_4bit.csv};
   \addlegendentry{\scriptsize MS, $(Q_{\text{ch}} = 4, Q_{\text{msg}} = 4)$}

   \addplot+[
			thick,
      mark=triangle*, 
      mark options={fill=green!60!black},
      color = green!60!black,
      ] 
      table [x index             = 0,
             y  index            = 1, 
             col sep=comma] {figure/fixed_5bit.csv};
   \addlegendentry{\scriptsize MS, $(Q_{\text{ch}} = 5, Q_{\text{msg}} = 5)$}

   \addplot+[
			thick,
      mark=x,
      mark options={fill=black},
      color = black,
      ] 
      table [x index             = 0,
             y  index            = 1, 
             col sep=comma] {figure/float_05iter.csv};
   \addlegendentry{\scriptsize MS, (floating-point)}

   \addplot+[
      mark=diamond*,
      mark options={fill=red!50!yellow},
      color = red!50!yellow,
      ]
      table [x index             = 0,
             y  index            = 1, 
             col sep=comma] {figure/float_05iter_OMS.csv};
   \addlegendentry{\scriptsize OMS, (Floating-point)$^2$}
   

\end{semilogyaxis}
\end{tikzpicture}
	\caption[footnote]{Frame error rate (FER) of the IEEE 802.3an LDPC code under floating-point MS decoding, fixed-point MS decoding with different bit-widths, LUT based decoding, and  floating-point \gls{OMS} decoding (offset=$0.5$) as reference, all with $I=5$ decoding iterations.}
	\label{fig:ecperf}	
\end{figure}
In Fig.~\ref{fig:ecperf}, we compare the performance of the IEEE 802.3an LDPC code under floating-point MS decoding, fixed-point MS decoding (with $Q_{\text{ch}} = Q_{\text{msg}} \in \{4,5\}$), and LUT-based decoding (with $Q_{\text{ch}} = 4$ and $Q_{\text{msg}} = 3$) when performing $I = 5$ decoding iterations with a flooding schedule. 
We also show the performance of a floating-point \glsfirst{OMS} decoder as a reference.\footnote{The reference simulation was obtained and matched with our simulation by using the open-source simulator provided by: Adrien Cassagne; Romain Tajan; Mathieu Léonardon; Baptiste Petit; Guillaume Delbergue; Thibaud Tonnellier; Camille Leroux; Olivier Hartmann, “AFF3CT: A Fast Forward Error Correction Tool,” 2016. [Online]. Available: https://doi.org/10.5281/zenodo.167837}
We observe that the fixed-point decoder with $Q_{\text{ch}} = Q_{\text{msg}} = 5$ has almost the same performance as the floating-point decoder, while the fixed-point decoder with $Q_{\text{ch}} = Q_{\text{msg}} = 4$ shows a significant loss with respect to the floating-point implementation. Thus, a standard \decMS would need to use at least $Q_{\text{ch}} = Q_{\text{msg}} = 5$ quantization bits. The LUT-based decoder, however, can match the performance of the floating-point decoder with only $Q_{\text{ch}} = 4$ channel quantization bits and $Q_{\text{msg}} = 3$ message quantization bits.\footnote{We note that reducing $Q_{\text{ch}}$ further results in a non-negligible loss with respect to the floating-point decoder.}
Additionally, with the above quantization bit choices, no noticeable error floor has been observed for both the \decMS and \decQN when $10^{10}$ frames have been transmitted (which corresponds to a BER $\approx 10^{-12}$ with the current block length). We note that, for the \decQN, the performance in the error floor region can be traded with the performance in the waterfall region by an appropriate choice of the design SNR for the LUTs~\cite{meidlinger2015asilomar}.

\subsection{LUT-Based Decoder Hardware Architecture}
The LUT-based serial message-transfer decoder hardware architecture is very similar to the \decMS architecture, described in Section~\ref{sec:Decoder1}. However, the LUT-based decoder can take advantage of the significantly fewer message bits that need to be transferred from one decoding stage to the next. This reduction reduces the number of \fastclk cycles per iteration, which in turn increases the throughput of the decoder according to \eqref{eq:fmax} provided that the CN/VN logic is sufficiently fast.
Moreover, the size of the buffers needed for the \gls{S/P} and \gls{P/S} conversions is also reduced significantly, which directly reduces the memory complexity of the decoder (see \eqref{eqn:finaltotalreg}).

On the negative side, we remark that the \gls{VN} units for each variable node stage (decoder iteration) of the \gls{LUT}-based decoder are different, which slightly complicates the hierarchical physical implementation as we will see later.
Furthermore, since $Q_{\text{ch}} > Q_{\text{msg}}$, we now need to transfer the channel \glspl{LLR} with multiple (two) bits per cycle to avoid the need to artificially limit the number of \fastclk cycles per iteration to $Q_{\text{ch}}$ rather than to the smaller  $Q_{\text{msg}}$. To reflect this modification, we redefine \eqref{eq:fmax} as $Q_{\text{max}}=\max (Q_{\text{msg}},\lceil \frac{Q_{\text{ch}}}{2} \rceil) $. 
While this partially parallel transfer of the channel \glspl{LLR} impacts routing congestion, we note that the overhead is negligible since the number of channel \glspl{LLR} is small compared to the total number of messages.

\section{Implementation}\label{sec:Implementation}
Despite the use of a serial message-transfer, the physical implementation of the decoders proposed in the previous sections requires special scrutiny since the number of global wires is still significant and the overall area is particularly large. 
Therefore, in this section, we propose and describe a \textit{pseudo-hierarchical} design methodology to implement the serial message-transfer architecture.

\subsection{Physical Design}
Due to the large number of identical blocks in the decoder architecture, a \textit{bottom-up} flow is expected to provide the best results.
The \gls{CN}, \gls{VN}, and \gls{DN} units are first placed and routed individually to build hard macros,\footnote{Note that for the \decQN there are different macros for each variable node stage as apposed to the \decMS.} and their timing and physical information are extracted.
These macros are then instantiated as large cells in the decoder top level.
We propose to treat the macros as \textit{custom standard-cells} with identical height to be able to perform the placement using the standard-cell placement, rather than the less capable macro placement of the backend tool, since in our case the number of hard macro instances is extremely large and the interconnect pattern is complex and highly irregular.

\begin{figure}%
	\centering
	\subfloat[][]{\includegraphics[scale=0.45]{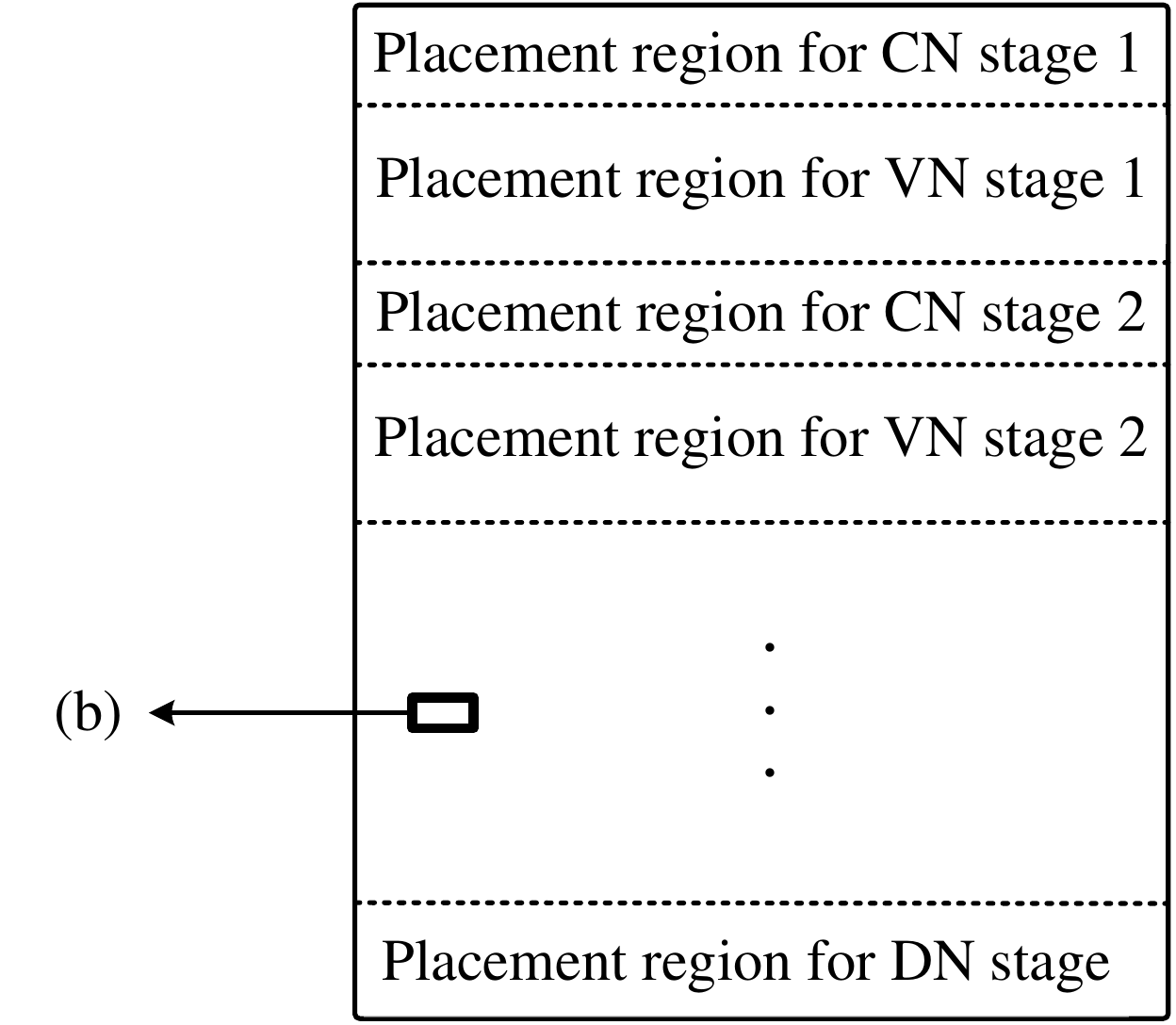} \label{subfig:floorplan} } \\ 
	\subfloat[][]{\includegraphics[scale=0.45]{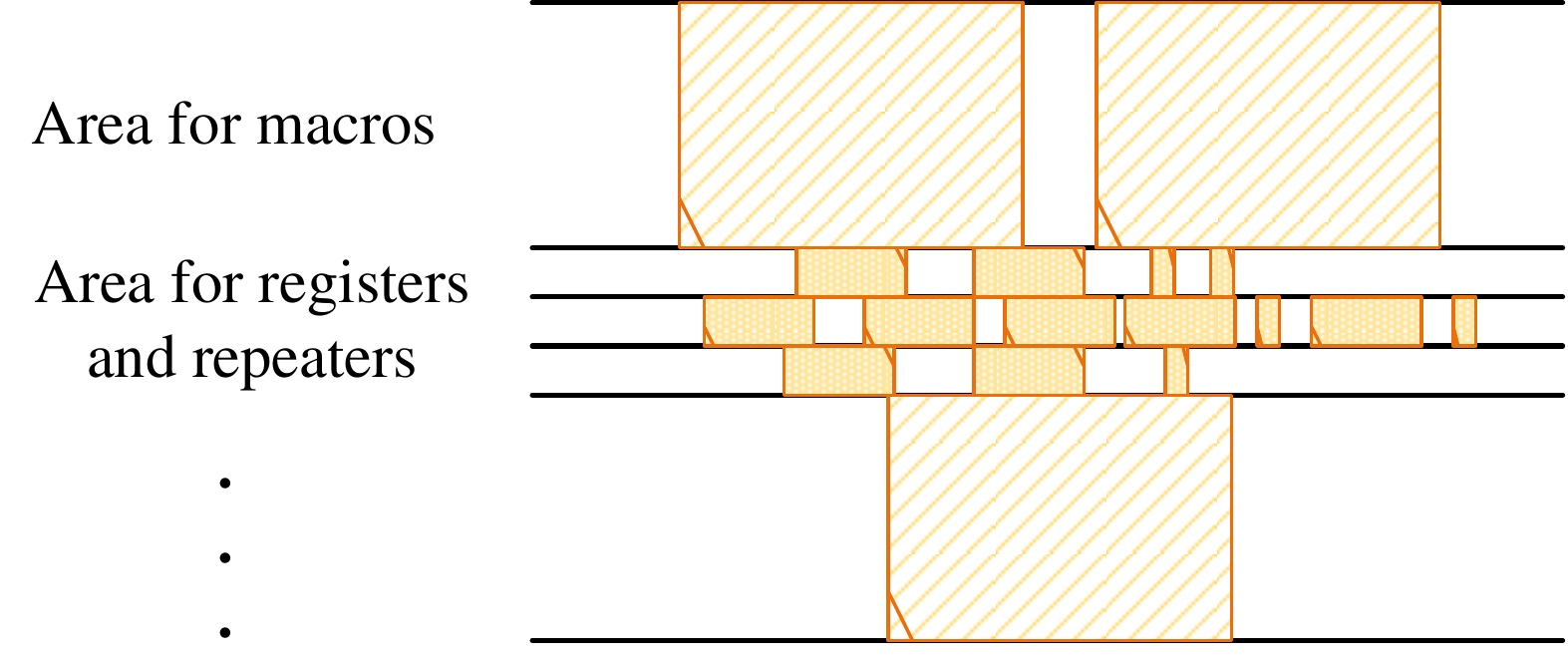} \label{subfig:rows} }
	\caption{
		The physical floorplan for serial message-transfer architecture,
		\protect\subref{subfig:floorplan} high level overview of the floorplan with dedicated placement regions for each decoder stage;
		and \protect\subref{subfig:rows} zoomed in overview showing rows structure for custom macros (large colored blocks) and conventional standard-cells (small colored blocks) placement.
	} 
	\label{fig:floorplan_rows}%
\end{figure}

Fig.~\ref{fig:floorplan_rows} illustrates the proposed physical floorplan for the decoders with the unrolled architecture.
In this floorplan, the \gls{CN} and \gls{VN} macros within each stage are constrained to be placed into dedicated regions (placement regions in Fig.~\ref{subfig:floorplan}).
This measure enforces the high-level structure of systolic array pipeline, but it also leaves freedom to the placement tool to choose the location for the macros in each stage to minimize routing congestion between stages.
Note that the linear floorplan has also the advantage of being scalable in the number of iterations since little interaction or interference exists between stages.
Furthermore, the \gls{CN} and \gls{VN} macros are placed in dedicated rows while the area between these rows is left for repeaters and for the register standard-cells for the channel \glspl{LLR} in the check node stages, as shown in Fig.~\ref{subfig:rows}. We note that the proposed floorplan and the encapsulation of the VN and CN macros as large standard-cells exploit the automated algorithm to optimize both custom and conventional standard-cells placement in order to alleviate the significant routing congestion.

\subsection{Timing and Area Optimization Flow}
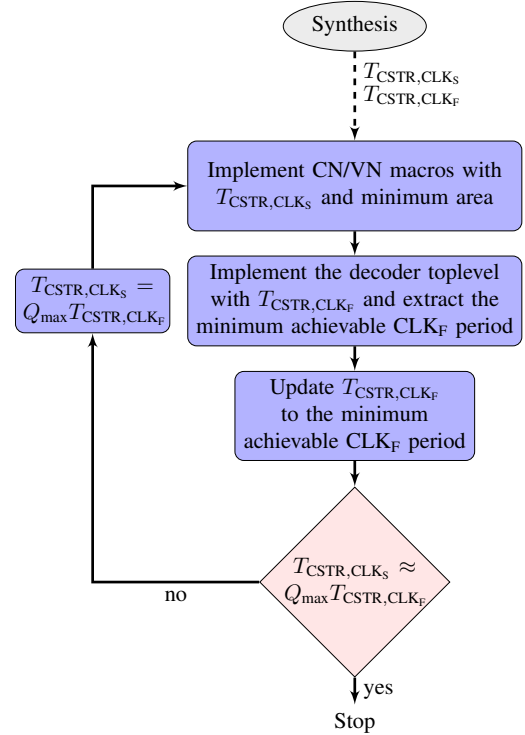
\begin{figure}
	\scalebox{0.85}{\tikzstyle{decision} = [diamond, draw, fill=red!10,
    text width=6.3em, text badly centered, node distance=2.6cm, inner sep=0pt]
    
\tikzstyle{block} = [rectangle, draw, fill=blue!30,
    text width=14.2em, text centered, rounded corners, minimum height=4em, node distance=2.5cm]		
 
\tikzstyle{smallblock} = [rectangle, draw, fill=blue!30,
    text width=10em, text centered, rounded corners, minimum height=2.6em, node distance=1.8cm]
  
\tikzstyle{smallblock2} = [rectangle, draw, fill=blue!30,
text width=6.2em, text centered, rounded corners, minimum height=2em, node distance=1.8cm]

\tikzstyle{cloud} = [draw, ellipse,fill=gray!15, node distance=2.5cm,
    minimum height=2em]
    
\tikzstyle{line} = [draw, very thick, color=black!100, -latex']

\begin{tikzpicture}[scale=2, node distance = 2.2cm, auto]       
    \node [cloud] (synthesis) {Synthesis};
    \node [block, below of=synthesis] (impl1) {Implement CN/VN macros with \TCmac and minimum area}; 
    \node [block, below of=impl1, node distance=1.8cm] (impl2) {Implement the decoder toplevel with \TCtop and extract the minimum achievable \fastclk period};
    \node [smallblock, below of=impl2]  (null1) {Update \TCtop to the minimum achievable \fastclk period};
    \node [smallblock2, left of=impl2, node distance=4.1cm]  (null2) {\TCmac$=$ \\ $Q_\text{max}$\TCtop};
    \node [decision, below of=null1] (decide) {\TCmac$\approx$ \\ $Q_\text{max}$\TCtop};
    \node [text centered, below of=decide] (stop) {Stop};
    \path [line, dashed] (synthesis) -- node [near start] {\TCmac} node [] {\TCtop}  (impl1) ;
    \path [line] (impl1) --  (impl2);
    \path [line] (impl2) -- (null1);
    \path [line] (null1) --  (decide);
    \path [line] (decide)  -| node [near start] {no} (null2);
    \path [line] (null2)  |- (impl1);
    \path [line] (decide) -- node [, color=black] {yes}(stop);

   \node (leyend) at (-0.5,0.6){
  \begin{tabular}{l@{: }l}
  \multicolumn{2}{c}{} \\
  \TCmac & timing constraint applied to \slowclk    \\
  \TCtop & timing constraint applied to \fastclk \\
  \end{tabular}
  };

\end{tikzpicture}}
	\caption{The proposed flowchart to optimize timing and area for the serial message-transfer architecture.}
	\label{fig:flowchart}
\end{figure}

Although the synthesis results can give an approximate evaluation for timing and area of the physical implementation, several iterations with different constraints are required to reach an optimal layout.
To this end, we propose the methodology illustrated in the flowchart of Fig.~\ref{fig:flowchart} to effectively implement the serial message-transfer architecture. The main idea behind this methodology is that three main factors directly contribute to the decoder throughput and also indirectly to the decoder area, as discussed in Section~\ref{sec:Decoder1} and specifically summarized in \eqref{eq:fmax}.
Our goal is to maximize the throughput at a minimum area.

We define the timing constraint applied to \slowclk as \TCmac, and the timing constrain applied to \fastclk as \TCtop.
The first step is to place and route the CN/VN macros based on \TCmac. This step is followed by the implementation of the decoder using \TCtop. (The initial constraints for the backend are thereby extracted from synthesis timing results.)
The fully placed and routed design can give an accurate routing delay, which will be used to update \TCtop and then \TCmac according to \eqref{eq:fmax}. 
The updated \TCmac will be used to re-implement the \gls{CN} and \gls{VN} macros within the minimum achievable area.

We note that for a long LDPC code with a large area and long routing delay (such as the one of the IEEE~802.3an standard), the first implementation starts with \mbox{\TCmac$<Q_\text{max}$\TCtop}.
After obtaining a realistic value for the \fastclk period (and hence for \TCtop) at the end of the implementation, the \TCmac will be updated to a larger value to approach \TCmac$\approx Q_\text{max}$\TCtop. Consequently, the \gls{CN} and \gls{VN} macro area and thus the decoder area will shrink in the second iteration, which result in larger achievable \fastclk frequency and hence smaller \TCtop and  \TCmac.
The feedback loop will reach the optimum point after a few iterations.

\begin{figure}
	\centering
	\subfloat[]{ \includegraphics[width=0.9in]{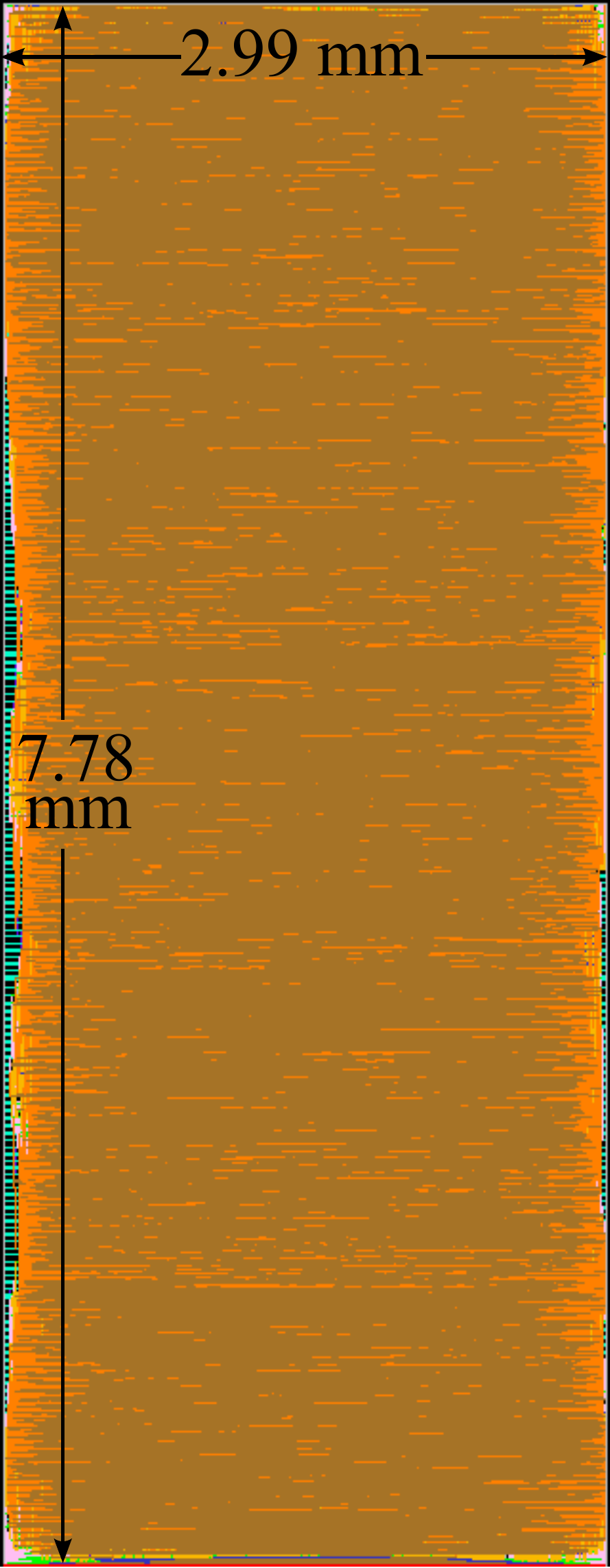} \label{subfig:ms} } \hspace{1cm}
	\subfloat[]{ \includegraphics[width=0.73in]{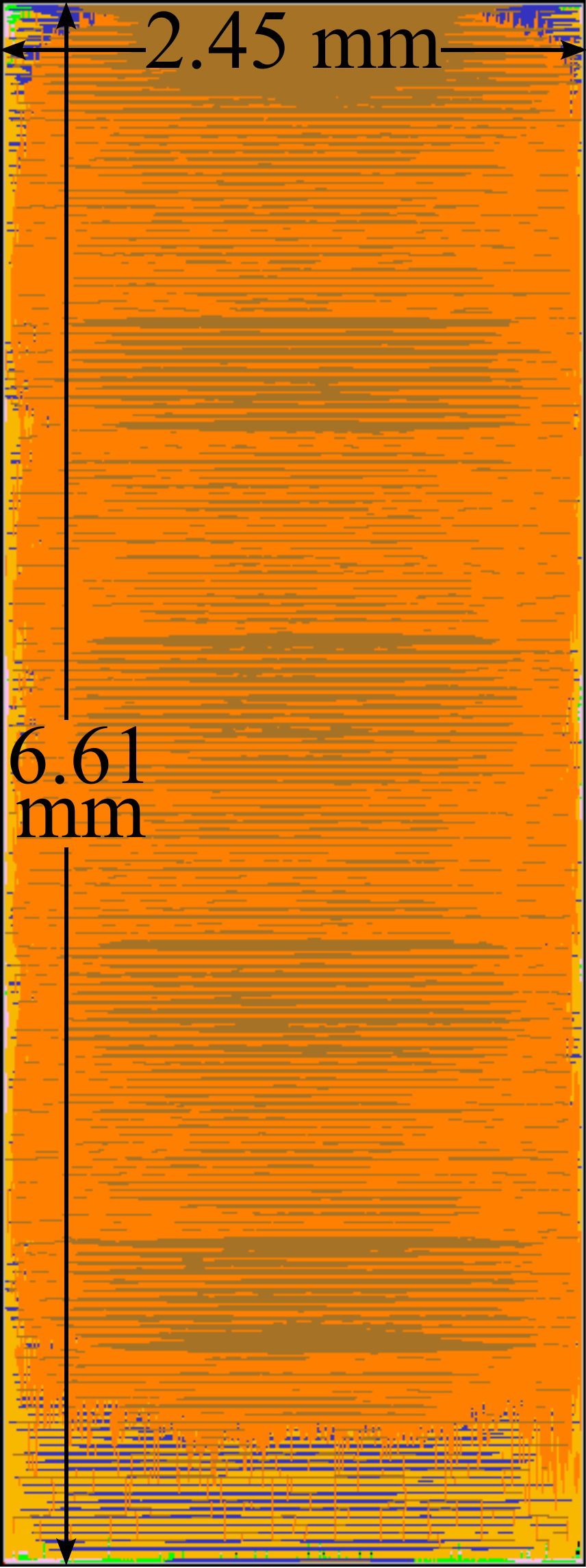} \label{subfig:lut} }
	\caption{Layouts for \protect\subref{subfig:ms} the \decMS and \protect\subref{subfig:lut} the \decQN.}
	\label{fig:layout} 
\end{figure}

\section{Results and Discussions}\label{sec:results}
To study the impact of the serial message-transfer architecture and the finite-alphabet decoding scheme, we have implemented the proposed architecture by employing the methodology explained in Section~\ref{sec:Implementation} and we analyzed the results for both \gls{MS} and LUT-based decoding. 
We used the parity check matrix of the LDPC code defined in the IEEE 802.3an standard~\cite{802.3an}, i.e., a $(2048, 1723)$ LDPC code of $R=0.8413$ with $d_v=6$ and $d_c=32$.
We used $I=5$ for both decoders and $Q_{\text{msg}}=Q_{\text{ch}}=5$ for the \decMS and $Q_{\text{msg}}=3$ and $Q_{\text{ch}}=4$ for the \decQN to achieve the same error-correction performance, as described in Section~\ref{sec:Decoder2}.
The decoders were synthesized from a VHDL description using Synopsys Design Compiler and placed and routed using Cadence Encounter Digital Implementation.
The layouts are shown in Fig.~\ref{fig:layout}.
The results are reported for a \mbox{$28\,$nm FD-SOI} library under typical operating conditions \mbox{($V_\mathrm{DD}=1\,$V, $T=25^{\circ}\,$C)}.

\begin{table}[t]
 	\caption{Critical path delays for \gls{MS} and \decQN}\label{tab:CriticalPath}
 	\centering
 	\begin{tabular}{c|cc}
 		\hline
 		\Xhline{2\arrayrulewidth}
 		 Path & \decMS & \decQN \\
 		\hline
 		\hline
 		CN [ns]    & $2.38$ & $1.42$ \\[1mm]		
 		VN [ns] & $0.96$ & $1.24$ \\[1mm]		
 		Routing [ns]  & $1.51$ & $1.16$ \\[1mm]		
 		\hline 
 	\end{tabular}
\end{table} 
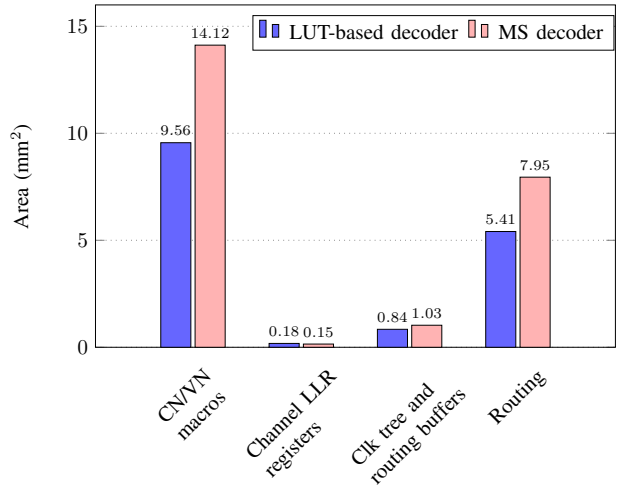
\begin{figure}[t]
	\scalebox{0.8}{\pgfplotstableread[col sep=comma]{
Name,A,B
CN/VN macros, 5398.7, 2391.7
Bla, 16196.1, 7175.1
Cla, 10153, 15583.4
Dla, 32081, 41902.6
}\datatable

\begin{tikzpicture}

\pgfplotsset{grid style={dotted,gray}}

\begin{axis}[
  ymajorgrids,
  ymin=0, ymax=16,
  ybar,
  bar width=0.5cm,
  ylabel=Area (mm$^2$),
  xtick=data,
  nodes near coords, every node near coord/.append style={font=\scriptsize},
  major x tick style = transparent,
  symbolic x coords={{CN/VN macros}, {Channel LLR reg.}, {Clk and routing buff.}, {Routing}},
  legend entries={\decQN, \decMS},
  legend style={at={(0.65,0.97)},
    anchor=north,
    legend columns=-1,
    column sep=1mm},
  x=1.8cm, enlarge x limits=0.3,
  x tick label style={rotate=45,align=left},
  xticklabels={
  CN/VN \\macros,
  Channel LLR \\registers,
  Clk tree and \\routing buffers,
  Routing
  },
]

\addplot [fill=blue!60] coordinates { ({CN/VN macros}, 9.56) ({Channel LLR reg.}, 0.18) ({Clk and routing buff.}, 0.84) ({Routing}, 5.41)};
\addplot [fill=red!30] coordinates { ({CN/VN macros}, 14.12) ({Channel LLR reg.}, 0.15) ({Clk and routing buff.}, 1.03) ({Routing}, 7.95)};

\end{axis}
\end{tikzpicture}} 
	\caption{Detailed area results for the LUT-based and \decMS with total area of $16.2\,$mm$^2$ and $23.3\,$mm$^2$, respectively.}
	\label{fig:area_utilization2} 
	\vspace{-4mm}
\end{figure}

\begin{table}[t]
 	\caption{Detailed area for CN/VN unit$^{\star}$}\label{tab:detailed_area}
 	\centering
 	\begin{tabular}{l|cc}
 		\hline
 		\Xhline{2\arrayrulewidth}
 		Component & \decMS & \decQN \\
 		\hline
 		\hline
 		CN unit logic [\si{\micro\metre}$^2$] & $1578$ & $485$ \\[1mm]
 		CN unit register [\si{\micro\metre}$^2$] & $1695$ & $971$ \\[1mm]
 		\bf{CN macro} [\si{\micro\metre}$^2$] & $\bf{3607}$ & $\bf{1510}$ \\[1mm]
 		\hline
 		VN unit logic [\si{\micro\metre}$^2$] & $315$ & $403^\dagger$ \\[1mm]	
 		VN unit register [\si{\micro\metre}$^2$] & $381$ & $235^\dagger$ \\[1mm]	
 		\bf{VN macro} [\si{\micro\metre}$^2$] & $\bf{755}$ & $\bf{646}^\dagger$ \\[1mm]	
 		\hline
 	\end{tabular} 
 	\begin{tablenotes}
 		\footnotesize 
 		\item \textit{$^\star$Logic and register areas are obtained form synthesis, and macro areas are the final post-layout results.}
 		\vspace{0.5mm}
 		\item \textit{$^\dagger$Although the VNs are different for each stage of the \decQN, their areas are similar and the result for one of them is reported here.}
 	\end{tablenotes}
\end{table} 

\begin{table}[t]
 	\caption{Power and energy efficiency comparison for \gls{MS} and \decQN}\label{tab:power}
 	\centering
 	\resizebox{\columnwidth}{!}{%
 		\begin{tabular}{c|cc}
 			\hline
 			\Xhline{2\arrayrulewidth}
 			& \decMS & \decQN \\
 			\hline
 			\hline
 			\multicolumn{1}{l|}{Total power @ $f_{\max}$ [mW]} & $12248$ & $13350$ \\[1mm]
 			\multicolumn{1}{l|}{Total power @ $662\,$MHz [mW]} & $12248$ & $10257$ \\[1mm]
 			\multicolumn{1}{l|}{Leakage power [mW]} & $7.44$ & $5.27$ \\[1mm]
 			\multicolumn{1}{l|}{Energy efficiency [pJ/bit]} & $45.2$ & $22.7$ \\[1mm]
 			\hline
 		\end{tabular}}
 		\vspace{-4mm}
\end{table}

\subsection{Delay Analysis}\label{subsec:DelayAnalysis}
In the serial message-transfer architecture, the critical path and, hence, the maximum decoding frequency are defined by~\eqref{eq:fmax}.
To investigate the impact of serially transferring the messages on the decoder throughput, we consider the delay of the following register-to-register critical paths for both the \gls{MS} and \decQN.
\subsubsection{CN Critical Path}
The \gls {CN} critical path ($T_{\text{CP},\text{CN}}$) is the path from the \gls{S/P} memory registers to the \gls{P/S} shift register within the \gls{CN} unit.
For both decoders, this path is essentially comprised of the logic cells for a sorter tree with a depth of four.

\subsubsection{VN Critical Path}
The \gls{VN} critical path ($T_{\text{CP},\text{VN}}$) is the path from \gls{S/P} memory registers to \gls{P/S} shift register within the \gls{VN} unit.
This path is dominated by an adder tree for the \decMS and an LUT tree for the \decQN.

\subsubsection{Routing Critical Path}
The routing critical path ($T_{\text{CP},\text{route}}$) comprises mainly the interconnect wires (and buffers) that connect the \gls{CN}/\gls{VN} unit \gls{S/P} shift register to the \gls{VN}/\gls{CN} unit \gls{P/S} shift register.

Table~\ref{tab:CriticalPath} summarizes the critical path delays of the \gls{CN}/\gls{VN} and the routing path.
Together with \eqref{eq:fmax}, the values in the table dictate the maximum achievable frequency for \slowclk and \fastclk, respectively, for both of the decoders with the proposed serial message-transfer architecture.
We note that the critical paths are reported after the timing and area constraints for the CN/VN macros and for the decoder toplevel are jointly optimized according to the flow shown in Fig.~\ref{fig:flowchart}.
According to \eqref{eq:fmax}, we observe that in both decoders the message transfer limits the slow clock \slowclk to a period of \mbox{$5\times1.51\,\text{ns}=7.55\,$ns} and \mbox{$3\times1.16\,\text{ns}=3.48\,$ns} for the \gls{MS} and the \decQN, respectively, where $1.51\,$ns and $1.16\,$ns are the corresponding minimum \fastclk periods.
Consequently, in our flow, the VN and CN units end up as optimized for minimum area only with relaxed and easy to meet timing constraints.

\begin{table*}[t]
	\caption{Implementation results for \gls{MS} and \decQN and comparison with other works}\label{tab:result_final} \vspace{-1mm}
	\centering
	\resizebox{\textwidth}{!}{%
		\begin{tabu}{l|cccccccc}
			\hline
			\Xhline{2\arrayrulewidth}
			& \decMS & \makecell{LUT-based\\decoder} & \cite{schlafer2013new} & \cite{zhang2010efficient} & \cite{mohsenin2010low} & \cite{cheng2014fully} & \cite{angarita2014reduced} & \cite{naderi2011delayed} \\
			\hline
			\hline
			\rule{0pt}{1\normalbaselineskip}
			Process technology & \multicolumn{2}{c}{$28\,$nm FD-SOI} & $65\,$nm CMOS & \makecell{$65\,$nm CMOS\\low-power} & $65\,$nm CMOS & $90\,$nm CMOS& $90\,$nm CMOS & $90\,$nm CMOS \\
			\;Supply voltage [V] &  \multicolumn{2}{c}{$1.0$} & $1.2$ & $1.2$ & $1.3$ & $0.9$ & $1.2$ & $1.0$\\ 
			\hline
			\rule{0pt}{1\normalbaselineskip}
			LDPC code  & \multicolumn{2}{c}{$(2048, 1723)$} & $(672, 546)$ & $(2048, 1723)$ & $(2048, 1723)$ & $(2048, 1723)$ & $(2048, 1723)$ & $(2048, 1723)$\\
			\;Node degree ($d_v, d_c$) & \multicolumn{2}{c}{$(6, 32)$} & $(3, 6)$ & $(6, 32)$ & $(6, 32)$ & $(6, 32)$ & $(6, 32)$ & $(6, 32)$\\
			\;Algorithm & min-sum & finite-alphabet & min-sum & \makecell{offset min-sum\\with post processor} & split-row & \makecell{normalized probablistic\\min-sum} & \makecell{reduced-complexity\\min-sum} & delayed stochastic \\
			\;$I_\text{max}$ & \multicolumn{2}{c}{$5$} & $9$ & $8$ & $11$ & $9$ & $30$ & $-$ \\
			\;Quantization bits & $5$ & $\bf{3}$ & $4$ & $4$ & $5$ & $4$ & $6$ & $5$ \\
			\;$E_b/N_0$ @ BER$=10^{-7}$ [dB] & $4.97$ & $4.95$ & $-$ & $4.25$ & $4.55$ & $4.4$ & $4.32$ & $4.7$\\
			\;Architecture & \multicolumn{2}{c}{unrolled full-parallel} & \makecell{unrolled\\full-parallel} & partial-parallel & full-parallel & full-parallel & full-parallel & full-parallel\\
			\hline
			\rule{0pt}{1\normalbaselineskip}
			Core area [mm$^2$] & $23.3$ & $16.2$ & $12.9$ & $5.05$ & $4.84$ & $9.6$ & $3.84$ & $3.93$ \\
			\;Area utilization [\%] & $66.4$ & $65.9$ & $76$ & $84.5$ & $97$ & $91$ & $-$ & $93$\\
			\;Max. frequency  ($f_{\max}$) [MHz] & $662$ & $862$ & $257$ & $700$ & $195$ & $199.6$ & $226$ & $750$\\
			\;Latency [ns] & $151$ &  $69.6$ & $105$ & $137$ & $56.4$ & $45.09$ & $-$ & $800$\\
			\;Throughput @ $I_\text{max}$ [Gbps]  & $271$ & $\bf{588}$ & $160.8$ & $\;13.3$ & $\;36.3$ & $45.42$ & $12.8$ & $172.4^\star$\\
			\;Power @ $f_{\max}$ [mW] & $12248$ & $13350$ & $5360$ & $2800$ & $1359$ & $1110$ & $1040$ & $-$\\
			\;Area eff. [Gbps/mm$^2$] & $11.6$ & $36.3$ & $12.5$ & $2.63$ & $7.5$ & $4.73$ & $3.34$ & $43.86$\\
			\;Energy per bit @ $I_\text{max}$ [pJ/bit] & $45.2$ & $22.7$ & $33.3$ & $\;\;210.5$ & $\;37.4$ & $24.44$ & $81.2$ & $-$\\
			\hline
			\rule{0pt}{1\normalbaselineskip}
			Scaled area eff.$^\dagger$ [Gbps/mm$^2$] & $11.6$ & $36.3$ & $156.4$ & $32.9$ & $93.8$ & $157.1$ & $110.9$ & $1456.8$\\
			\;Scaled energy per bit$^\ddagger$ [pJ/bit] & $45.2$ & $22.7$ & $10$ & $63$ & $9.5$ & $9.4$ & $17.5$ & $-$\\
			\hline
		\end{tabu}
	}
	\begin{tablenotes}
		\footnotesize 
		\item \textit{$^\star$Maximum throughput @ $E_b/N_0=5.5\,\mathrm{dB}$ (Note that throughput @ $I_\text{max}$ is not reported in the original paper)}
		\vspace{0.5mm}
		\item \textit{$^\dagger$Scaling is done by $S^3$ where $S$ is the relative dimension to $28$ (Note that this is very rough and optimistic since it does not apply to the interconnects)}
		\item \textit{$^\ddagger$Scaling is done by $1/SU^2$ where $U$ is the relative voltage to $1.0$}
	\end{tablenotes}
	\vspace{-4mm}
\end{table*} 
\vspace{-4mm}
\subsection{Area Analysis}
Fig.~\ref{fig:area_utilization2} illustrates the area distribution among the various components after the layout. 
The area utilization is approximately $67$\% for both decoders. While almost $62$\% of the layout is filled with \gls{CN}/\gls{VN} macros and registers, the clock tree and routing buffers occupy around $5$\%.
Furthermore, we see a $44$\% difference in total area between the decoders due to the fact that the total area for \gls{CN} and \gls{VN} macros is $14.12\,$mm$^2$ in the \gls{MS} decoder, as opposed to only $9.56\,$mm$^2$ in the \decQN.

To understand this fact, we list the area of each \gls{CN}/\gls{VN} macro in Table~\ref{tab:detailed_area}.
According to this table, the finite-alphabet message passing algorithm leads to significantly smaller \gls{CN} processors because of two important factors:
first, the bit-width reduction of the messages directly affects the data-path area, and second, the quantized messages in the \decQN are processed directly in the sorter tree of the \glspl{CN} without the need to compute their absolute values. 
However, \gls{VN} processors are less area-efficient in the \decQN in comparison with the ones of the \decMS.
This is caused by the fact that the LUT-based computations are, in general, less area-efficient than the conventional arithmetic~based update rules.
Thus, the logic area of the \gls{VN} in the \decQN is larger, even though their input/output bit-width is smaller.
Another contributing factor in the Table~\ref{tab:detailed_area} is the register area, which is defined by the number of \gls{S/P} and \gls{P/S} registers.
For those, the $40$\% reduction of bit-width in the \decQN is directly noticeable in the register area for both \gls{CN} and \gls{VN} units.
Altogether, the \gls{CN} and \gls{VN} macros in the \decQN are $58$\% and $14$\% smaller, respectively, compared to those of the \decMS.

\subsection{Power Analysis}
The energy which is consumed by each decoder is proportional to the capacitance, which in turn is related to the decoder area.
Also, the number of required \fastclk cycles for the serial message-transfer to decode one codeword, which is inversely proportional to the decoding throughput at a constant frequency, directly contributes to the consumed energy for each decoded bit.
Therefore, we analyze both the total power and the energy efficiency of the decoders using post-layout vector-based power analysis.\footnote{We first extract the parasitic information of both the hard macros and the top level from the placement and routing tool and then read and link them using a power computation tool to generate the complete parasitic information.}
The results are reported in Table~\ref{tab:power}.
We note that the total powers are calculated at $f_{\text{max}}$ for both decoders. Also, for comparison purpose, we have calculated the total powers at a constant \fastclk frequency, here $\min (f_{\text{max},\text{MS}}, f_{\text{max},\text{LUT}})=662\,$MHz, for both decoders and note them in the Table~\ref{tab:power}.
According to this table, the total power consumption of the \decQN is $16.2$\% smaller than that of the \decMS.
Furthermore, by considering the fact that the \decQN has $66.7$\% higher throughput than the \decMS at a similar \fastclk frequency, the energy efficiency of the \decQN is almost $2$ times better in comparison with the \decMS.

\subsection{Summary and Final Comparison}
The final post-layout results for our \gls{MS} and LUT-based decoders and also for some other recently implemented decoders are summarized in Table~\ref{tab:result_final}.
Our \decQN runs at a maximum \fastclk frequency of $f_{\text{max},\text{LUT}}=862\,$MHz and delivers a sustained throughput of $588\,$Gbps, while it occupies $16.2\,$mm$^2$ area and dissipates $22.7\,$pJ/bit.
Compared to the \decMS, the \decQN is $1.4$$\times$ smaller, $2.2$$\times$ faster, and thus $3.1$$\times$ more area efficient.
It also has $16.2$\% lower power dissipation and $2$$\times$ better energy efficiency, when the decoding throughout is taken into account.

The work in \cite{schlafer2013new} is the only other unrolled full-parallel decoder in literature, but it is designed for the IEEE~802.11ad~\cite{802.11ad} code, which has a shorter block length and smaller node degrees ($d_v=3$ and $d_c=6$ as opposed to $d_v=6$ and $d_c=32$ for the code used in the design reported in this paper).
The work of \cite{zhang2010efficient}, \cite{mohsenin2010low,naderi2011delayed,cheng2014fully}, and \cite{angarita2014reduced} are for the same IEEE~802.3an code considered in this paper, but with partial-parallel and full-parallel architectures.
The proposed LUT-based decoder has more than an order of magnitude higher throughput compared to \cite{mohsenin2010low} and \cite{cheng2014fully}, and  three times higher throughput compared to \cite{naderi2011delayed}, while the maximum throughput of the proposed decoder is maintained for all SNR scenarios as it does not require early termination to achieve a high throughput.
The area efficiency of the proposed unrolled full-parallel architecture, however, is inferior to the one of the decoders in \cite{mohsenin2010low,cheng2014fully} and \cite{angarita2014reduced} with full-parallel architecture due to the repeated routing overhead between the decoder stages in our design.

\section{Conclusion}\label{sec:conclusion}

An ultra high throughput LDPC decoder with a serial message-transfer architecture and based on non-uniform quantization of messages was proposed to achieve the highest decoding throughput in literature.
The proposed decoder architecture is an unrolled full-parallel architecture with serialized messages for \gls{CN}/\gls{VN} units, which was enabled by employing \gls{S/P} and \gls{P/S} shift registers at the inputs and outputs of each unit.
The proposed quantized message passing algorithm replaces conventional \gls{MS}, resulting in $40$\% reduction in message bit-width without any performance penalty.
This algorithm was implemented by using generic LUTs instead of adders for \glspl{VN} while the \glspl{CN} remained unchanged compared to \gls{MS} decoding.
Placement and routing results in $28\,$nm \mbox{FD-SOI} show that the LUT-based serial message-transfer decoder delivers $0.588\,$Tbps throughput and is $3.1$ times more area efficient and $2$ times more energy efficient in comparison with the MS decoder with serial message-transfer architecture.

\section*{Acknowledgment}
This work was supported by  the Swiss National Science Foundation (SNSF) under the project number 200021-153640.

\bibliographystyle{IEEEtran}
\vspace{-5 pt}
\bibliography{IEEEabrv,./share/LDPC}

\begin{IEEEbiography}	
	[{\includegraphics[width=1in,height=1.25in,clip,keepaspectratio,angle=0]{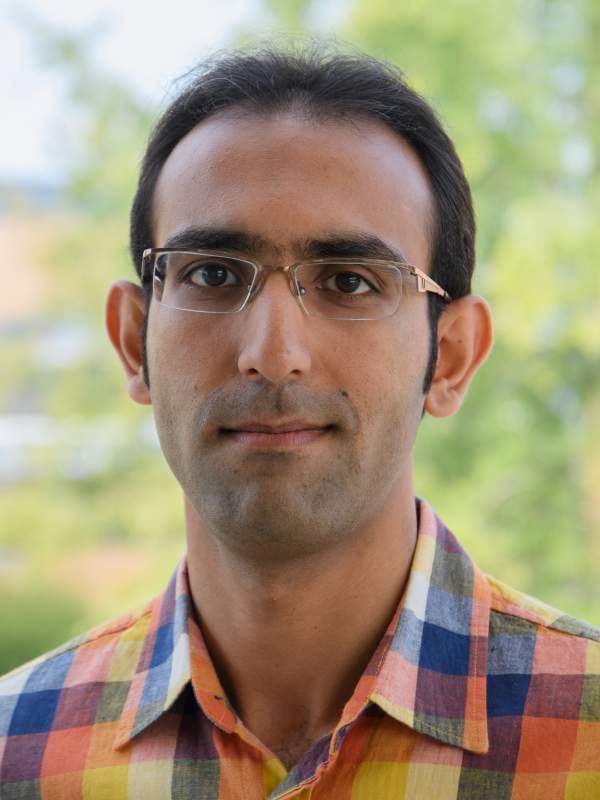}}]{Reza Ghanaatian}
was born in Jahrom, Iran, in 1988. He received the M.Sc. degree in digital systems from the Department of Electrical Engineering, Sharif University of Technology (SUT), Tehran, Iran, in 2012. He was with Advanced Integrated Circuit Design Laboratory (AICDL), at SUT from 2011 to 2013, working on field-programmable gate array based systems for wireless and optical communication applications.

In 2014, Mr. Ghanaatian joined Telecommunications Circuits Laboratory (TCL) at EPFL, Lausanne, Switzerland, working towards his Ph.D. His current research interests include VLSI circuits for signal processing and communications as well as approximate computing techniques for energy efficient system design.
\end{IEEEbiography}
\vspace{-1cm}
\begin{IEEEbiography}
	[{\includegraphics[width=1in,height=1.25in,clip,keepaspectratio]{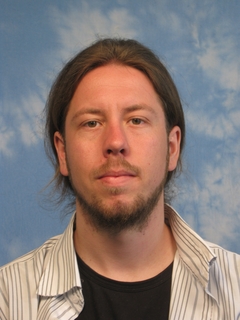}}]{Alexios Balatsoukas-Stimming}
	received the Diploma and M.Sc. degrees in Electronics and Computer Engineering from the Technical University of Crete, Greece, in 2010 and 2012, respectively. His M.Sc. studies were supported by a scholarship from the Alexander S. Onassis foundation. He received the Ph.D degree in Computer and Communications Sciences from EPFL, Switzerland, where he performed his research at the Telecommunications Circuits Laboratory. He serves as reviewer for several IEEE Journals and Conferences, and has been recognized as an Exemplary Reviewer by the IEEE Wireless Communications Letters in 2013 and 2014. His current research interests include VLSI circuits for signal processing and communications, as well as error correction coding theory and practice.
\end{IEEEbiography}
\vspace{-1cm}
\begin{IEEEbiography}
	    [{\includegraphics[width=1in,height=1.25in,clip,keepaspectratio]{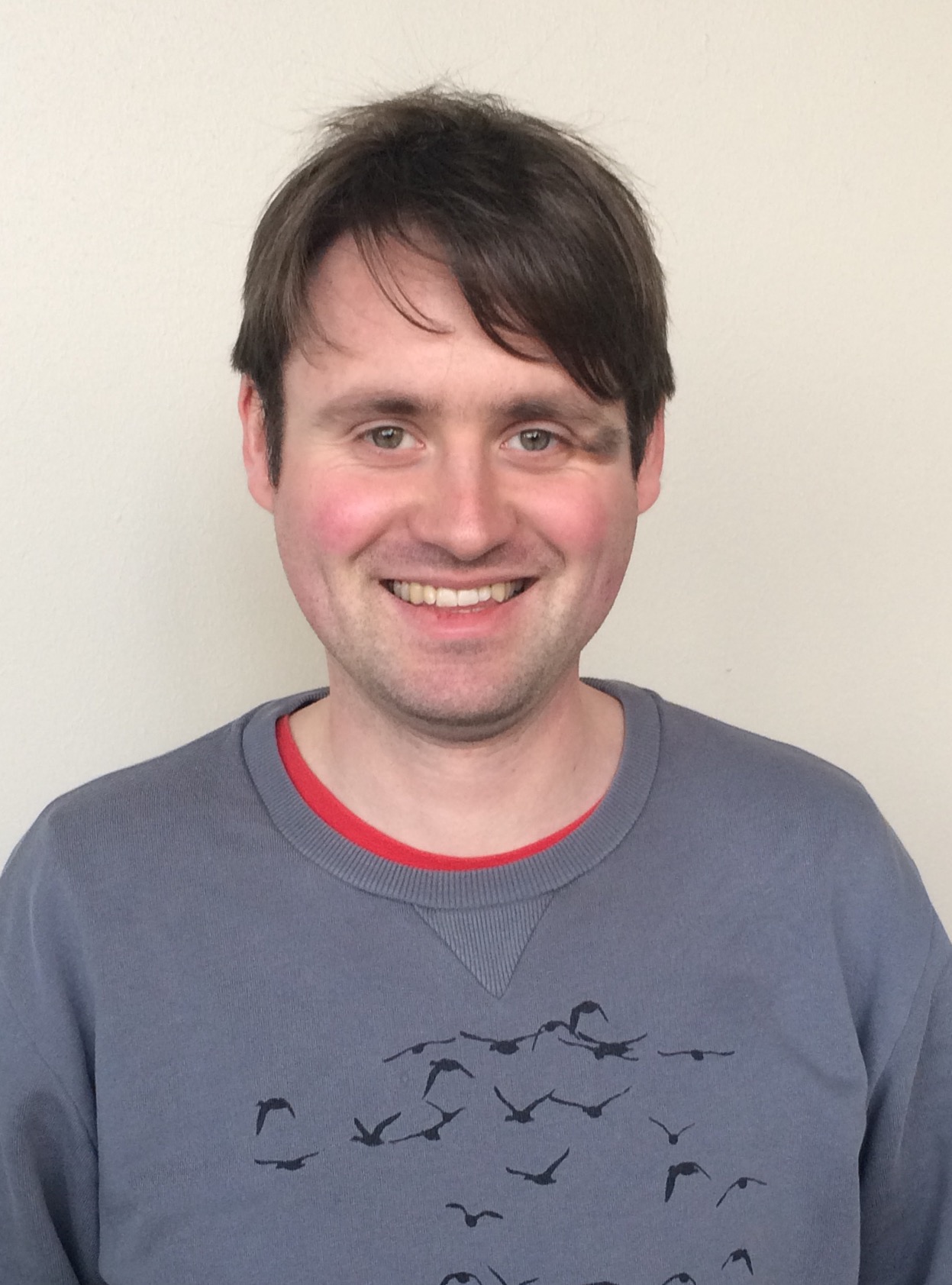}}]{Thomas Christoph M\"{u}ller}
	    (S'16) received the bachelor's degree in information technology from the Schmalkalden University of Applied Sciences, Schmalkalden, Germany, in 2011, and the master's degree in System-on-Chip from Lund University, Sweden, in 2013. He has been working as Project/Research Assistant at Lund University and at Technical University of Denmark, Kongens Lyngby, Denmark. Currently he is pursuing his Ph.D. degree in the Telecommunication Circuits Laboratory (TCL) at the \'Ecole polytechnique f\'ed\'erale de Lausanne (EPFL), Switzerland with a research focus on digital implementation and, an emphasis on low power, variation mitigation and design methodology.
\end{IEEEbiography}
\vspace{-1cm}
\begin{IEEEbiography}
	[{\includegraphics[width=1in,height=1.25in,clip,keepaspectratio]{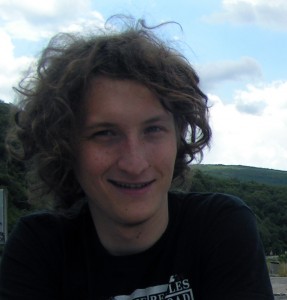}}]{Michael Meidlinger}
was born in Austria in 1989. He received his Bsc. and Msc. degrees from Technische Universität (TU) Wien, Vienna, Austria in 2011 and 2013 respectively, both with distinction.
From 2011 to 2013, Michael has been working in the field of mobile communication research and helped to develop the Vienna LTE-A Simulators.
Since 2013, Michael is part of the Communication Theory group at TU Wien, working towards his  Ph.D. His current research interests include quantizer design
for telecommunication receivers as well as error correction coding and superposition modulation techniques.
\end{IEEEbiography}

\vspace{-1cm}
\begin{IEEEbiography}
	[{\includegraphics[width=1in,height=1.25in,clip,keepaspectratio]{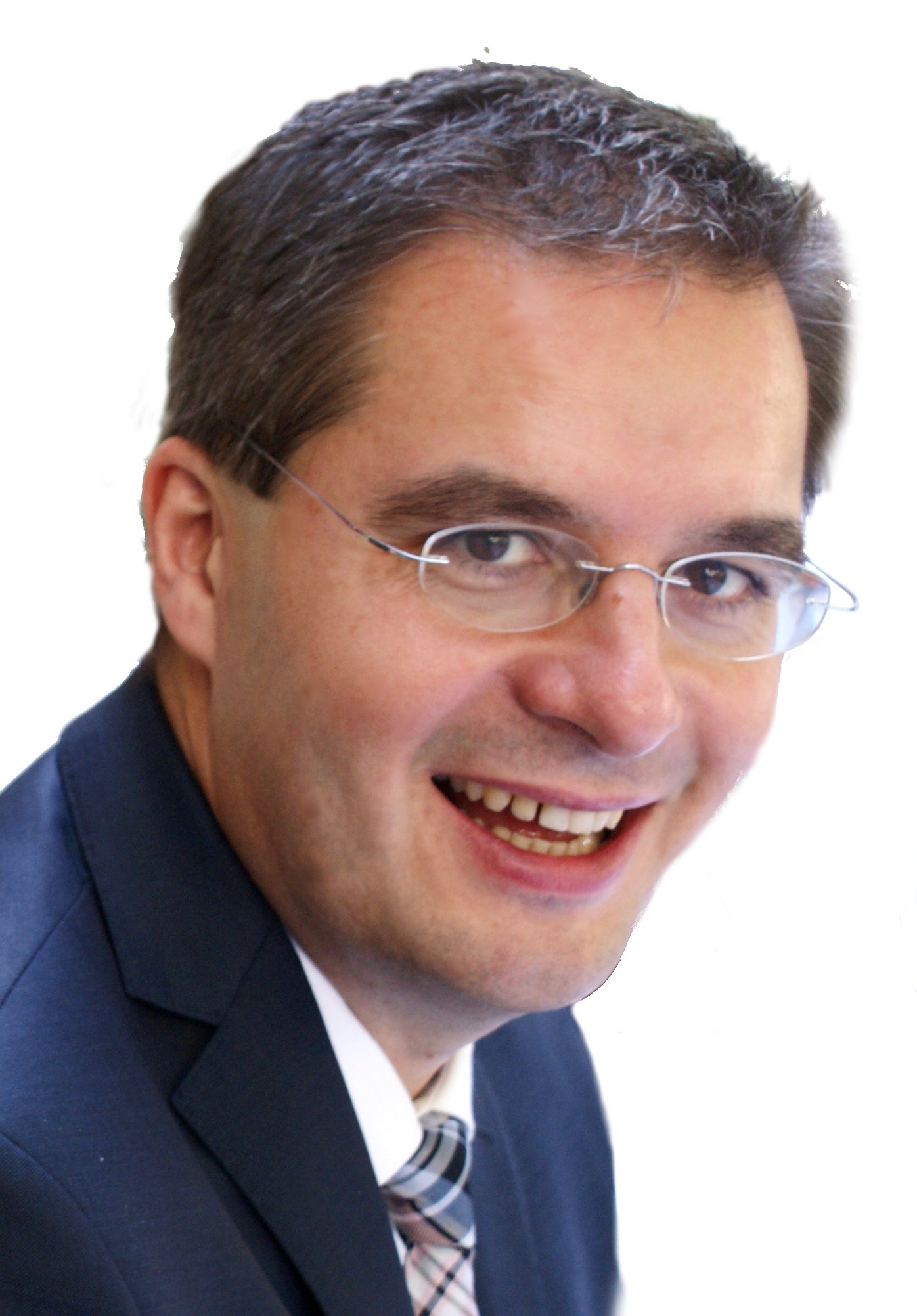}}]{Gerald Matz}
	received the Dipl.-Ing.\ (1994) and Dr.\ techn.\ (2000) degrees in Electrical Engineering and the Habilitation degree (2004) for Communication Systems from Vienna University of Technology, Austria. He currently holds a tenured Associate Professor position
	with the Institute of Telecommunications, Vienna University of
	Technology. He has held visiting positions with the Laboratoire des Signaux et Syst\`{e}mes at Ecole Sup\'{e}rieure d’Electricit\'{e} (France, 2004),
	the Communication Theory Lab at ETH Zurich (Switzerland, 2007), and with Ecole Nationale Sup\'{e}rieure d'Electrotechnique, d'Electronique, d'Informatique et d'Hydraulique de Toulouse (France, 2011).
	
	Prof.~Matz has directed or actively participated in several research projects funded by the Austrian Science Fund (FWF), by the Viennese Science and Technology Fund (WWTF), and by the European Union. He has published some 200 scientific articles in international journals, conference proceedings, and edited books. He is co-editor of the book Wireless Communications over Rapidly Time-Varying Channels (New York: Academic, 2011). His research interests include wireless networks, statistical signal processing, information theory, and big data.
	
	Prof.~Matz served as as a member of the IEEE SPS Technical Committee on Signal Processing Theory and Methods and of the IEEE SPS Technical Committee on Signal Processing for Communications and Networking. He was an Associate Editor of the IEEE Transactions on Information Theory
	(2013-2015), of the IEEE Transactions on Signal Processing (2006–2010), of the EURASIP Journal Signal Processing (2007–2010), and of the IEEE Signal Processing Letters (2004–2008). He was Technical Program Chair of Asilomar 2016, Technical Program Co-Chair of EUSIPCO 2004, Technical
	Area Chair for “MIMO Communications and Signal Processing” at Asilomar
	2012, and Technical Area Chair for “Array Processing” at Asilomar 2015.
	He has been a member of the Technical Program Committee of numerous international conferences. In 2006 he received the Kardinal Innitzer Most Promising Young Investigator Award. He is an IEEE Senior Member and a member of the \"{O}VE.
\end{IEEEbiography}
\vspace{-1cm}
\begin{IEEEbiography}
    [{\includegraphics[width=1in,height=1.25in,clip,keepaspectratio]{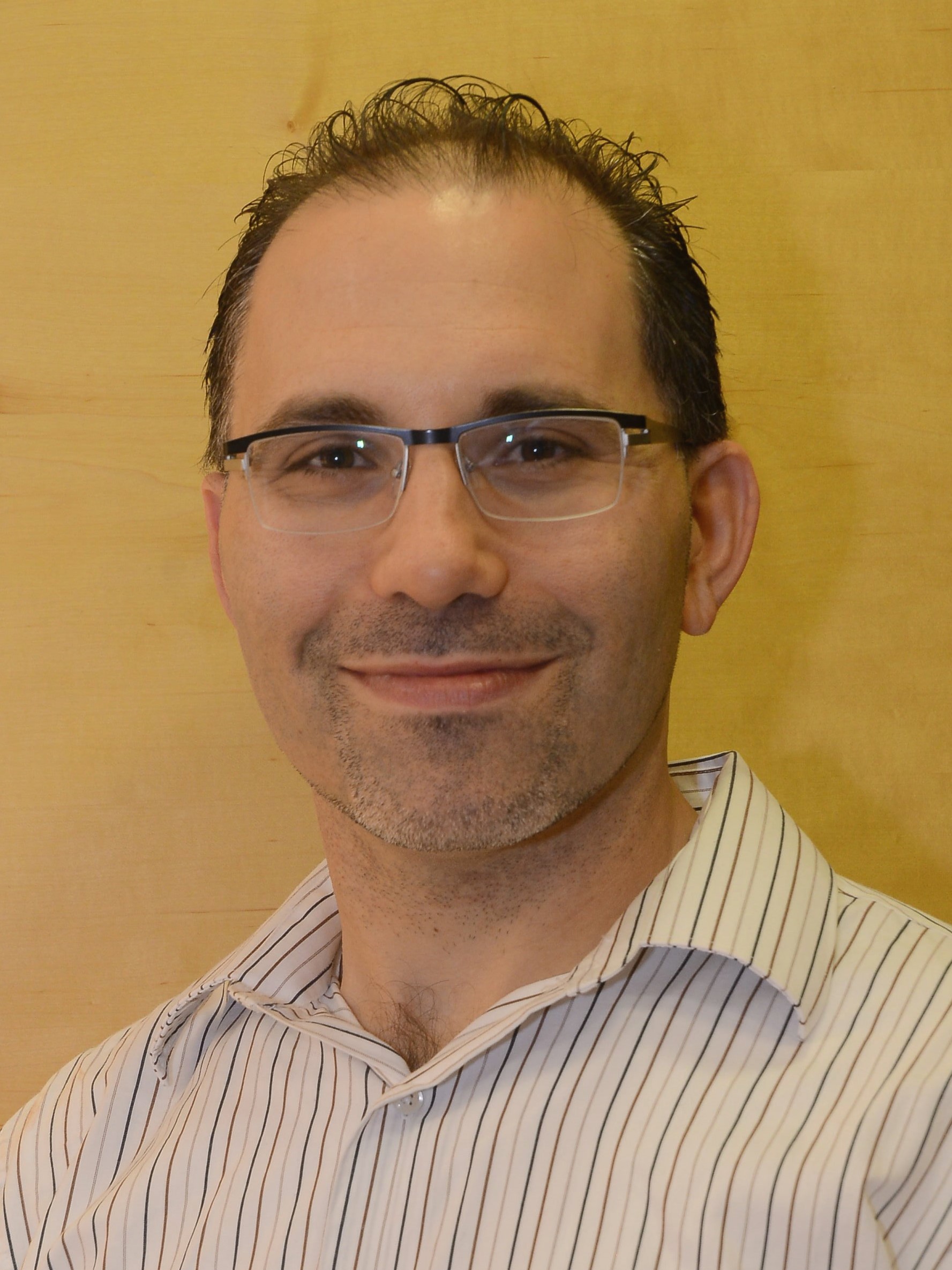}}]{Adam Teman}
	received the Ph.D., M.Sc., and B.Sc. degrees in Electrical Engineering from Ben-Gurion University (BGU), Be'er Sheva, Israel in 2006, 2011, and 2014, respectively.
	He worked as a Design Engineer at Marvell Semiconductors from 2006 to 2007, with an emphasis on Physical Implementation. 
	Dr. Teman's research interests include low-voltage digital design, energy efficient SRAM, NVM, and eDRAM memory arrays, low power CMOS image sensors, low power design techniques for digital and analog VLSI chips, significance-driven approximate computing, and process tolerant design techniques. He has authored more than 40 scientific papers and 4 patent applications and is an associate editor at the Microelectronics Journal and a technical committee member of several IEEE conferences. In 2010--2012, Dr. Teman was honored with the Electrical Engineering Department's Teaching Excellence recognition at BGU, and in 2011, he was awarded with BGU's Outstanding Project award. Dr. Teman received the Yizhak Ben-Ya’akov HaCohen Prize in 2010, the BGU Rector's Prize for Outstanding Academic Achievement in 2012, the Wolf Foundation Scholarship for excellence of 2012 and the Intel Prize for Ph.D. students in 2013. His doctoral studies were conducted under a Kreitman Foundation Fellowship.
	Dr. Teman was a post-doctoral researcher at the Telecommunications Circuits Lab (TCL) at the \'Ecole Polytechnique F\'ed\'erale de Lausanne (EPFL), Switzerland under a Swiss Government Excellence Scholarship from 2014--2015. 
	In October 2015, Dr. Teman joined the faculty of engineering at Bar-Ilan University, Ramat Gan, Israel in 2015, where he is currently a tenure track researcher in the department of electrical engineering and a leading member of the Emerging Nanoscaled Integrated Circuits and Systems (EnICS) Labs.
	
\end{IEEEbiography}
\vspace{-9mm}
\begin{IEEEbiography}
	[{\includegraphics[width=1in,height=1.25in,clip,keepaspectratio]{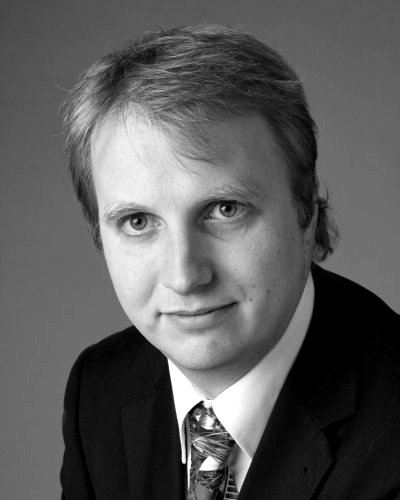}}]{Andreas Burg}
 (S'97-M'05) was born in Munich, Germany, in 1975. He received his Dipl.-Ing. degree from the Swiss Federal Institute of Technology (ETH) Zurich, Zurich, Switzerland, in 2000, and the Dr. sc. techn. degree from the Integrated Systems Laboratory of ETH Zurich, in 2006. In 1998, he worked at Siemens Semiconductors, San Jose, CA. During his doctoral studies, he worked at Bell Labs Wireless Research for a total of one year. From 2006 to 2007, he was a postdoctoral researcher at the Integrated Systems Laboratory and at the Communication Theory Group of the ETH Zurich. In 2007 he co-founded Celestrius, an ETH-spinoff in the field of MIMO wireless communication, where he was responsible for the ASIC development as Director for VLSI. In January 2009, he joined ETH Zurich as SNF Assistant Professor and as head of the Signal Processing Circuits and Systems group at the Integrated Systems Laboratory. Since January 2011, he has been a Tenure Track Assistant Professor at the Ecole Polytechnique Federale de Lausanne (EPFL) where he is leading the Telecommunications Circuits Laboratory. 

In 2000, Mr. Burg received the “Willi Studer Award” and the ETH Medal for his diploma and his diploma thesis, respectively. Mr. Burg was also awarded an ETH Medal for his Ph.D. dissertation in 2006. In 2008, he received a 4-years grant from the Swiss National Science Foundation (SNF) for an SNF Assistant Professorship. With his students he received the best paper award from the EURASIP Journal on Image and Video Processing in 2013 and best demo/paper awards at ISCAS 2013, ICECS 2013, and at ACSSC 2007. 

He has served on the TPC of various conferences on signal processing, communications, and VLSI. He was a TPC co-chair for VLSI-SoC 2012 and the TCP co-chair for ESSCIRC 2016 and SiPS 2017. He served as an Editor for the IEEE Transaction of Circuits and Systems in 2013 and is on the Editorial board of the Springer Microelectronics Journal and the MDPI Journal on Low Power Electronics and its Applications. He is also a member of the EURASIP SAT SPCN and of the IEEE TC-DISPS. 
\end{IEEEbiography}
\vfill

\end{document}